\documentstyle[12pt,aaspp4]{article}
\newcommand{\etal}{\hbox{ et~al.}}

\def\PsfigVersion{1.10}
\def\setDriver{\DvipsDriver} 
\ifx\undefined\psfig\else \fi
\let\LaTeXAtSign=\@
\let\@=\relax
\edef\psfigRestoreAt{\catcode`\@=\number\catcode`@\relax}
\catcode`\@=11\relax
\newwrite\@unused
\def\ps@typeout#1{{\let\protect\string\immediate\write\@unused{#1}}}

\def\DvipsDriver{
	\ps@typeout{psfig/tex \PsfigVersion -dvips}
\def\PsfigSpecials{\DvipsSpecials} 	\def\ps@dir{/}
\def\ps@predir{} }
\def\OzTeXDriver{
	\ps@typeout{psfig/tex \PsfigVersion -oztex}
	\def\PsfigSpecials{\OzTeXSpecials}
	\def\ps@dir{:}
	\def\ps@predir{:}
	\catcode`\^^J=5
}

\def\figurepath{./:}

\def\DoPaths#1{\expandafter\EachPath#1\stoplist}
\def\leer{}
\def\EachPath#1:#2\stoplist{
  \ExistsFile{#1}{\SearchedFile}
  \ifx#2\leer
  \else
    \expandafter\EachPath#2\stoplist
  \fi}
\def\ps@dir{/}
\def\ExistsFile#1#2{%
   \openin1=\ps@predir#1\ps@dir#2
   \ifeof1
       \closein1
   \else
       \closein1
        \ifx\ps@founddir\leer
           \edef\ps@founddir{#1}
        \fi
   \fi}
\def\get@dir#1{%
  \def\ps@founddir{}
  \def\SearchedFile{#1}
  \DoPaths\figurepath
}

\def\@nnil{\@nil}
\def\@empty{}
\def\@psdonoop#1\@@#2#3{}
\def\@psdo#1:=#2\do#3{\edef\@psdotmp{#2}\ifx\@psdotmp\@empty \else
    \expandafter\@psdoloop#2,\@nil,\@nil\@@#1{#3}\fi}
\def\@psdoloop#1,#2,#3\@@#4#5{\def#4{#1}\ifx #4\@nnil \else
       #5\def#4{#2}\ifx #4\@nnil \else#5\@ipsdoloop #3\@@#4{#5}\fi\fi}
\def\@ipsdoloop#1,#2\@@#3#4{\def#3{#1}\ifx #3\@nnil 
       \let\@nextwhile=\@psdonoop \else
      #4\relax\let\@nextwhile=\@ipsdoloop\fi\@nextwhile#2\@@#3{#4}}
\def\@tpsdo#1:=#2\do#3{\xdef\@psdotmp{#2}\ifx\@psdotmp\@empty \else
    \@tpsdoloop#2\@nil\@nil\@@#1{#3}\fi}
\def\@tpsdoloop#1#2\@@#3#4{\def#3{#1}\ifx #3\@nnil 
       \let\@nextwhile=\@psdonoop \else
      #4\relax\let\@nextwhile=\@tpsdoloop\fi\@nextwhile#2\@@#3{#4}}
\ifx\undefined\fbox
\newdimen\fboxrule
\newdimen\fboxsep
\newdimen\ps@tempdima
\newbox\ps@tempboxa
\long\def\fbox#1{\leavevmode\setbox\ps@tempboxa\hbox{#1}\ps@tempdima\fboxrule
    \advance\ps@tempdima \fboxsep \advance\ps@tempdima \dp\ps@tempboxa
   \hbox{\lower \ps@tempdima\hbox
  {\vbox{\hrule height \fboxrule
          \hbox{\vrule width \fboxrule \hskip\fboxsep
          \vbox{\vskip\fboxsep \box\ps@tempboxa\vskip\fboxsep}\hskip 
                 \fboxsep\vrule width \fboxrule}
                 \hrule height \fboxrule}}}}
\fi
\newread\ps@stream
\newif\ifnot@eof       
\newif\if@noisy        
\newif\if@atend        
\newif\if@psfile       
{\catcode`\%=12\global\gdef\epsf@start{
\def\epsf@PS{PS}
\def\epsf@getbb#1{%
\openin\ps@stream=\ps@predir#1
\ifeof\ps@stream\ps@typeout{Error, File #1 not found}\else
   {\not@eoftrue \chardef\other=12
    \def\do##1{\catcode`##1=\other}\dospecials \catcode`\ =10
    \loop
       \if@psfile
	  \read\ps@stream to \epsf@fileline
       \else{
	  \obeyspaces
          \read\ps@stream to \epsf@tmp\global\let\epsf@fileline\epsf@tmp}
       \fi
       \ifeof\ps@stream\not@eoffalse\else
       \if@psfile\else
       \expandafter\epsf@test\epsf@fileline:. \\%
       \fi
          \expandafter\epsf@aux\epsf@fileline:. \\%
       \fi
   \ifnot@eof\repeat
   }\closein\ps@stream\fi}%
\long\def\epsf@test#1#2#3:#4\\{\def\epsf@testit{#1#2}
			\ifx\epsf@testit\epsf@start\else
\ps@typeout{Warning! File does not start with `\epsf@start'.  It may not be a PostScript file.}
			\fi
			\@psfiletrue} 
{\catcode`\%=12\global\let\epsf@percent=
\long\def\epsf@aux#1#2:#3\\{\ifx#1\epsf@percent
   \def\epsf@testit{#2}\ifx\epsf@testit\epsf@bblit
	\@atendfalse
        \epsf@atend #3 . \\%
	\if@atend	
	   \if@verbose{
		\ps@typeout{psfig: found `(atend)'; continuing search}
	   }\fi
        \else
        \epsf@grab #3 . . . \\%
        \not@eoffalse
        \global\no@bbfalse
        \fi
   \fi\fi}%
\def\epsf@grab #1 #2 #3 #4 #5\\{%
   \global\def\epsf@llx{#1}\ifx\epsf@llx\empty
      \epsf@grab #2 #3 #4 #5 .\\\else
   \global\def\epsf@lly{#2}%
   \global\def\epsf@urx{#3}\global\def\epsf@ury{#4}\fi}%
\def\epsf@atendlit{(atend)} 
\def\epsf@atend #1 #2 #3\\{%
   \def\epsf@tmp{#1}\ifx\epsf@tmp\empty
      \epsf@atend #2 #3 .\\\else
   \ifx\epsf@tmp\epsf@atendlit\@atendtrue\fi\fi}

\chardef\psletter = 11 
\chardef\other = 12

\newif \ifdebug 
\newif\ifc@mpute 
\c@mputetrue 

\let\then = \relax
\def\r@dian{pt }
\let\r@dians = \r@dian
\let\dimensionless@nit = \r@dian
\let\dimensionless@nits = \dimensionless@nit
\def\internal@nit{sp }
\let\internal@nits = \internal@nit
\newif\ifstillc@nverging
\def \Mess@ge #1{\ifdebug \then \message {#1} \fi}

{ 
	\catcode `\@ = \psletter
	\gdef \nodimen {\expandafter \n@dimen \the \dimen}
	\gdef \term #1 #2 #3%
	       {\edef \t@ {\the #1}
		\edef \t@@ {\expandafter \n@dimen \the #2\r@dian}%
		\t@rm {\t@} {\t@@} {#3}%
	       }
	\gdef \t@rm #1 #2 #3%
	       {{%
		\count 0 = 0
		\dimen 0 = 1 \dimensionless@nit
		\dimen 2 = #2\relax
		\Mess@ge {Calculating term #1 of \nodimen 2}%
		\loop
		\ifnum	\count 0 < #1
		\then	\advance \count 0 by 1
			\Mess@ge {Iteration \the \count 0 \space}%
			\Multiply \dimen 0 by {\dimen 2}%
			\Mess@ge {After multiplication, term = \nodimen 0}%
			\Divide \dimen 0 by {\count 0}%
			\Mess@ge {After division, term = \nodimen 0}%
		\repeat
		\Mess@ge {Final value for term #1 of 
				\nodimen 2 \space is \nodimen 0}%
		\xdef \Term {#3 = \nodimen 0 \r@dians}%
		\aftergroup \Term
	       }}
	\catcode `\p = \other
	\catcode `\t = \other
	\gdef \n@dimen #1pt{#1} 
}

\def \Divide #1by #2{\divide #1 by #2} 

\def \Multiply #1by #2
       {{
	\count 0 = #1\relax
	\count 2 = #2\relax
	\count 4 = 65536
	\Mess@ge {Before scaling, count 0 = \the \count 0 \space and
			count 2 = \the \count 2}%
	\ifnum	\count 0 > 32767 
	\then	\divide \count 0 by 4
		\divide \count 4 by 4
	\else	\ifnum	\count 0 < -32767
		\then	\divide \count 0 by 4
			\divide \count 4 by 4
		\else
		\fi
	\fi
	\ifnum	\count 2 > 32767 
	\then	\divide \count 2 by 4
		\divide \count 4 by 4
	\else	\ifnum	\count 2 < -32767
		\then	\divide \count 2 by 4
			\divide \count 4 by 4
		\else
		\fi
	\fi
	\multiply \count 0 by \count 2
	\divide \count 0 by \count 4
	\xdef \product {#1 = \the \count 0 \internal@nits}%
	\aftergroup \product
       }}

\def\r@duce{\ifdim\dimen0 > 90\r@dian \then   
		\multiply\dimen0 by -1
		\advance\dimen0 by 180\r@dian
		\r@duce
	    \else \ifdim\dimen0 < -90\r@dian \then  
		\advance\dimen0 by 360\r@dian
		\r@duce
		\fi
	    \fi}

\def\Sine#1%
       {{%
	\dimen 0 = #1 \r@dian
	\r@duce
	\ifdim\dimen0 = -90\r@dian \then
	   \dimen4 = -1\r@dian
	   \c@mputefalse
	\fi
	\ifdim\dimen0 = 90\r@dian \then
	   \dimen4 = 1\r@dian
	   \c@mputefalse
	\fi
	\ifdim\dimen0 = 0\r@dian \then
	   \dimen4 = 0\r@dian
	   \c@mputefalse
	\fi
	\ifc@mpute \then
		\divide\dimen0 by 180
		\dimen0=3.141592654\dimen0
		\dimen 2 = 3.1415926535897963\r@dian 
		\divide\dimen 2 by 2 
		\Mess@ge {Sin: calculating Sin of \nodimen 0}%
		\count 0 = 1 
		\dimen 2 = 1 \r@dian 
		\dimen 4 = 0 \r@dian 
		\loop
			\ifnum	\dimen 2 = 0 
			\then	\stillc@nvergingfalse 
			\else	\stillc@nvergingtrue
			\fi
			\ifstillc@nverging 
			\then	\term {\count 0} {\dimen 0} {\dimen 2}%
				\advance \count 0 by 2
				\count 2 = \count 0
				\divide \count 2 by 2
				\ifodd	\count 2 
				\then	\advance \dimen 4 by \dimen 2
				\else	\advance \dimen 4 by -\dimen 2
				\fi
		\repeat
	\fi		
			\xdef \sine {\nodimen 4}%
       }}

\def\Cosine#1{\ifx\sine\UnDefined\edef\Savesine{\relax}\else
		             \edef\Savesine{\sine}\fi
	{\dimen0=#1\r@dian\advance\dimen0 by 90\r@dian
	 \Sine{\nodimen 0}
	 \xdef\cosine{\sine}
	 \xdef\sine{\Savesine}}}	      

\def\psdraft{
	\def\@psdraft{0}
}
\def\psfull{
	\def\@psdraft{100}
}

\psfull

\newif\if@scalefirst
\def\psscalefirst{\@scalefirsttrue}
\def\psrotatefirst{\@scalefirstfalse}
\psrotatefirst

\newif\if@draftbox
\def\psnodraftbox{
	\@draftboxfalse
}
\def\psdraftbox{
	\@draftboxtrue
}
\@draftboxtrue

\newif\if@prologfile
\newif\if@postlogfile
\def\pssilent{
	\@noisyfalse
}
\def\psnoisy{
	\@noisytrue
}
\psnoisy
\newif\if@bbllx
\newif\if@bblly
\newif\if@bburx
\newif\if@bbury
\newif\if@height
\newif\if@width
\newif\if@rheight
\newif\if@rwidth
\newif\if@angle
\newif\if@clip
\newif\if@verbose
\def\@p@@sclip#1{\@cliptrue}
\newif\if@decmpr
\def\@p@@sfigure#1{\def\@p@sfile{null}\def\@p@sbbfile{null}\@decmprfalse
   \openin1=\ps@predir#1
   \ifeof1
	\closein1
	\get@dir{#1}
	\ifx\ps@founddir\leer
		\openin1=\ps@predir#1.bb
		\ifeof1
			\closein1
			\get@dir{#1.bb}
			\ifx\ps@founddir\leer
				\ps@typeout{Can't find #1 in \figurepath}
			\else
				\@decmprtrue
				\def\@p@sfile{\ps@founddir\ps@dir#1}
				\def\@p@sbbfile{\ps@founddir\ps@dir#1.bb}
			\fi
		\else
			\closein1
			\@decmprtrue
			\def\@p@sfile{#1}
			\def\@p@sbbfile{#1.bb}
		\fi
	\else
		\def\@p@sfile{\ps@founddir\ps@dir#1}
		\def\@p@sbbfile{\ps@founddir\ps@dir#1}
	\fi
   \else
	\closein1
	\def\@p@sfile{#1}
	\def\@p@sbbfile{#1}
   \fi
}
\def\@p@@sfile#1{\@p@@sfigure{#1}}
\def\@p@@sbbllx#1{
		\@bbllxtrue
		\dimen100=#1
		\edef\@p@sbbllx{\number\dimen100}
}
\def\@p@@sbblly#1{
		\@bbllytrue
		\dimen100=#1
		\edef\@p@sbblly{\number\dimen100}
}
\def\@p@@sbburx#1{
		\@bburxtrue
		\dimen100=#1
		\edef\@p@sbburx{\number\dimen100}
}
\def\@p@@sbbury#1{
		\@bburytrue
		\dimen100=#1
		\edef\@p@sbbury{\number\dimen100}
}
\def\@p@@sheight#1{
		\@heighttrue
		\dimen100=#1
   		\edef\@p@sheight{\number\dimen100}
}
\def\@p@@swidth#1{
		\@widthtrue
		\dimen100=#1
		\edef\@p@swidth{\number\dimen100}
}
\def\@p@@srheight#1{
		\@rheighttrue
		\dimen100=#1
		\edef\@p@srheight{\number\dimen100}
}
\def\@p@@srwidth#1{
		\@rwidthtrue
		\dimen100=#1
		\edef\@p@srwidth{\number\dimen100}
}
\def\@p@@sangle#1{
		\@angletrue
		\edef\@p@sangle{#1} 
}
\def\@p@@ssilent#1{ 
		\@verbosefalse
}
\def\@p@@sprolog#1{\@prologfiletrue\def\@prologfileval{#1}}
\def\@p@@spostlog#1{\@postlogfiletrue\def\@postlogfileval{#1}}
\def\@cs@name#1{\csname #1\endcsname}
\def\@setparms#1=#2,{\@cs@name{@p@@s#1}{#2}}
\def\ps@init@parms{
		\@bbllxfalse \@bbllyfalse
		\@bburxfalse \@bburyfalse
		\@heightfalse \@widthfalse
		\@rheightfalse \@rwidthfalse
		\def\@p@sbbllx{}\def\@p@sbblly{}
		\def\@p@sbburx{}\def\@p@sbbury{}
		\def\@p@sheight{}\def\@p@swidth{}
		\def\@p@srheight{}\def\@p@srwidth{}
		\def\@p@sangle{0}
		\def\@p@sfile{} \def\@p@sbbfile{}
		\def\@p@scost{10}
		\def\@sc{}
		\@prologfilefalse
		\@postlogfilefalse
		\@clipfalse
		\if@noisy
			\@verbosetrue
		\else
			\@verbosefalse
		\fi
}
\def\parse@ps@parms#1{
	 	\@psdo\@psfiga:=#1\do
		   {\expandafter\@setparms\@psfiga,}}
\newif\ifno@bb
\def\bb@missing{
	\if@verbose{
		\ps@typeout{psfig: searching \@p@sbbfile \space  for bounding box}
	}\fi
	\no@bbtrue
	\epsf@getbb{\@p@sbbfile}
        \ifno@bb \else \bb@cull\epsf@llx\epsf@lly\epsf@urx\epsf@ury\fi
}	
\def\bb@cull#1#2#3#4{
	\dimen100=#1 bp\edef\@p@sbbllx{\number\dimen100}
	\dimen100=#2 bp\edef\@p@sbblly{\number\dimen100}
	\dimen100=#3 bp\edef\@p@sbburx{\number\dimen100}
	\dimen100=#4 bp\edef\@p@sbbury{\number\dimen100}
	\no@bbfalse
}
\newdimen\p@intvaluex
\newdimen\p@intvaluey
\def\rotate@#1#2{{\dimen0=#1 sp\dimen1=#2 sp
		  \global\p@intvaluex=\cosine\dimen0
		  \dimen3=\sine\dimen1
		  \global\advance\p@intvaluex by -\dimen3
		  \global\p@intvaluey=\sine\dimen0
		  \dimen3=\cosine\dimen1
		  \global\advance\p@intvaluey by \dimen3
		  }}
\def\compute@bb{
		\no@bbfalse
		\if@bbllx \else \no@bbtrue \fi
		\if@bblly \else \no@bbtrue \fi
		\if@bburx \else \no@bbtrue \fi
		\if@bbury \else \no@bbtrue \fi
		\ifno@bb \bb@missing \fi
		\ifno@bb \ps@typeout{FATAL ERROR: no bb supplied or found}
			\no-bb-error
		\fi
		\count203=\@p@sbburx
		\count204=\@p@sbbury
		\advance\count203 by -\@p@sbbllx
		\advance\count204 by -\@p@sbblly
		\edef\ps@bbw{\number\count203}
		\edef\ps@bbh{\number\count204}
		\if@angle 
			\Sine{\@p@sangle}\Cosine{\@p@sangle}
	        	{\dimen100=\maxdimen\xdef\r@p@sbbllx{\number\dimen100}
					    \xdef\r@p@sbblly{\number\dimen100}
			                    \xdef\r@p@sbburx{-\number\dimen100}
					    \xdef\r@p@sbbury{-\number\dimen100}}
                        \def\minmaxtest{
			   \ifnum\number\p@intvaluex<\r@p@sbbllx
			      \xdef\r@p@sbbllx{\number\p@intvaluex}\fi
			   \ifnum\number\p@intvaluex>\r@p@sbburx
			      \xdef\r@p@sbburx{\number\p@intvaluex}\fi
			   \ifnum\number\p@intvaluey<\r@p@sbblly
			      \xdef\r@p@sbblly{\number\p@intvaluey}\fi
			   \ifnum\number\p@intvaluey>\r@p@sbbury
			      \xdef\r@p@sbbury{\number\p@intvaluey}\fi
			   }
			\rotate@{\@p@sbbllx}{\@p@sbblly}
			\minmaxtest
			\rotate@{\@p@sbbllx}{\@p@sbbury}
			\minmaxtest
			\rotate@{\@p@sbburx}{\@p@sbblly}
			\minmaxtest
			\rotate@{\@p@sbburx}{\@p@sbbury}
			\minmaxtest
			\edef\@p@sbbllx{\r@p@sbbllx}\edef\@p@sbblly{\r@p@sbblly}
			\edef\@p@sbburx{\r@p@sbburx}\edef\@p@sbbury{\r@p@sbbury}
		\fi
		\count203=\@p@sbburx
		\count204=\@p@sbbury
		\advance\count203 by -\@p@sbbllx
		\advance\count204 by -\@p@sbblly
		\edef\@bbw{\number\count203}
		\edef\@bbh{\number\count204}
}
\def\in@hundreds#1#2#3{\count240=#2 \count241=#3
		     \count100=\count240	
		     \divide\count100 by \count241
		     \count101=\count100
		     \multiply\count101 by \count241
		     \advance\count240 by -\count101
		     \multiply\count240 by 10
		     \count101=\count240	
		     \divide\count101 by \count241
		     \count102=\count101
		     \multiply\count102 by \count241
		     \advance\count240 by -\count102
		     \multiply\count240 by 10
		     \count102=\count240	
		     \divide\count102 by \count241
		     \count200=#1\count205=0
		     \count201=\count200
			\multiply\count201 by \count100
		 	\advance\count205 by \count201
		     \count201=\count200
			\divide\count201 by 10
			\multiply\count201 by \count101
			\advance\count205 by \count201
		     \count201=\count200
			\divide\count201 by 100
			\multiply\count201 by \count102
			\advance\count205 by \count201
		     \edef\@result{\number\count205}
}
\def\compute@wfromh{
		\in@hundreds{\@p@sheight}{\@bbw}{\@bbh}
		\edef\@p@swidth{\@result}
}
\def\compute@hfromw{
	        \in@hundreds{\@p@swidth}{\@bbh}{\@bbw}
		\edef\@p@sheight{\@result}
}
\def\compute@handw{
		\if@height 
			\if@width
			\else
				\compute@wfromh
			\fi
		\else 
			\if@width
				\compute@hfromw
			\else
				\edef\@p@sheight{\@bbh}
				\edef\@p@swidth{\@bbw}
			\fi
		\fi
}
\def\compute@resv{
		\if@rheight \else \edef\@p@srheight{\@p@sheight} \fi
		\if@rwidth \else \edef\@p@srwidth{\@p@swidth} \fi
}
\def\compute@sizes{
	\compute@bb
	\if@scalefirst\if@angle
	\if@width
	   \in@hundreds{\@p@swidth}{\@bbw}{\ps@bbw}
	   \edef\@p@swidth{\@result}
	\fi
	\if@height
	   \in@hundreds{\@p@sheight}{\@bbh}{\ps@bbh}
	   \edef\@p@sheight{\@result}
	\fi
	\fi\fi
	\compute@handw
	\compute@resv}
\def\OzTeXSpecials{
	\special{empty.ps /@isp {true} def}
	\special{empty.ps \@p@swidth \space \@p@sheight \space
			\@p@sbbllx \space \@p@sbblly \space
			\@p@sbburx \space \@p@sbbury \space
			startTexFig \space }
	\if@clip{
		\if@verbose{
			\ps@typeout{(clip)}
		}\fi
		\special{empty.ps doclip \space }
	}\fi
	\if@angle{
		\if@verbose{
			\ps@typeout{(rotate)}
		}\fi
		\special {empty.ps \@p@sangle \space rotate \space} 
	}\fi
	\if@prologfile
	    \special{\@prologfileval \space } \fi
	\if@decmpr{
		\if@verbose{
			\ps@typeout{psfig: Compression not available
			in OzTeX version \space }
		}\fi
	}\else{
		\if@verbose{
			\ps@typeout{psfig: including \@p@sfile \space }
		}\fi
		\special{epsf=\ps@predir\@p@sfile \space }
	}\fi
	\if@postlogfile
	    \special{\@postlogfileval \space } \fi
	\special{empty.ps /@isp {false} def}
}
\def\DvipsSpecials{
	\special{ps::[begin] 	\@p@swidth \space \@p@sheight \space
			\@p@sbbllx \space \@p@sbblly \space
			\@p@sbburx \space \@p@sbbury \space
			startTexFig \space }
	\if@clip{
		\if@verbose{
			\ps@typeout{(clip)}
		}\fi
		\special{ps:: doclip \space }
	}\fi
	\if@angle
		\if@verbose{
			\ps@typeout{(clip)}
		}\fi
		\special {ps:: \@p@sangle \space rotate \space} 
	\fi
	\if@prologfile
	    \special{ps: plotfile \@prologfileval \space } \fi
	\if@decmpr{
		\if@verbose{
			\ps@typeout{psfig: including \@p@sfile.Z \space }
		}\fi
		\special{ps: plotfile "`zcat \@p@sfile.Z" \space }
	}\else{
		\if@verbose{
			\ps@typeout{psfig: including \@p@sfile \space }
		}\fi
		\special{ps: plotfile \@p@sfile \space }
	}\fi
	\if@postlogfile
	    \special{ps: plotfile \@postlogfileval \space } \fi
	\special{ps::[end] endTexFig \space }
}
\def\psfig#1{\vbox {
	\ps@init@parms
	\parse@ps@parms{#1}
	\compute@sizes
	\ifnum\@p@scost<\@psdraft{
		\PsfigSpecials 
		\vbox to \@p@srheight sp{
			\hbox to \@p@srwidth sp{
				\hss
			}
		\vss
		}
	}\else{
		\if@draftbox{		
			\hbox{\fbox{\vbox to \@p@srheight sp{
			\vss
			\hbox to \@p@srwidth sp{ \hss 
			 \hss }
			\vss
			}}}
		}\else{
			\vbox to \@p@srheight sp{
			\vss
			\hbox to \@p@srwidth sp{\hss}
			\vss
			}
		}\fi

	}\fi
}}
\psfigRestoreAt
\setDriver
\let\@=\LaTeXAtSign

\begin{document}

\title {A Study of Gravitational Lens Chromaticity with the Hubble Space Telescope\footnote{
     Based on observations made with the NASA/ESA Hubble Space Telescope. 
     The Space Telescope Science Institute is operated by the Association of
     Universities for Research in Astronomy, Inc. under NASA contract 
     NAS 5-26555.}}


\author{J.A. Mu\~noz$^1$, E. Mediavilla$^2$,  C.S. Kochanek$^3$, E.E. Falco$^4$ and A.M. Mosquera $^{1,3}$} 

\bigskip
\affil{$^{1}$Departamento de Astronom\'{\i}a y Astrof\'{\i}sica, Universidad
     de Valencia, E-46100 Burjassot, Valencia, Spain}
\affil{$^{2}$Instituto de Astrof\'{\i}sica de Canarias, E-38200 La Laguna,
     Tenerife, Spain}
\affil{$^{3}$Department of Astronomy, The Ohio State University, Columbus,
 OH 43210, USA}
\affil{$^{4}$Harvard-Smithsonian Center for Astrophysics, 60 Garden St.,
Cambridge, MA 02138, USA}
\affil{email: jmunoz@uv.es}


\begin{abstract}

We report Hubble Space Telescope observations of 6 gravitational lenses with the Advanced Camera for Surveys. We measured
the flux ratios between the lensed images in 6 filters from 8140\AA\ to 2200\AA.
In 3 of the systems, HE0512$-$3329, B1600+434, and H1413$+$117,
we were able to construct UV extinction curves partially overlapping the 2175\AA\ feature 
and characterize the properties of the dust relative to the Galaxy and the
Magellanic Clouds.  In HE1104$-$1804 we  detect chromatic microlensing and use it to
study the physical properties of the quasar accretion disk. For a Gaussian 
model of the disk  $\exp(-r^2/2 r_s^2)$, scaling with wavelength as $r_s \propto \lambda^p$,
we estimate 
$r_s( \lambda3363)=4^{+4}_{-2}$ ($7\pm 4$)  light-days and  $p=1.1\pm 0.6$ ($1.0\pm 0.6$) for a logarithmic 
(linear) prior on $r_s$. 
The remaining two systems, FBQ0951+2635 and SBS1520+530,
yielded no useful estimates of extinction or chromatic microlensing.
\end{abstract}


\section{Introduction}

The wavelength-dependent flux ratios between gravitationally lensed images of
quasars can be used to probe both the extinction law in the lens galaxy and
the unresolved structure of the quasar.  Each image is differentially extincted
by dust along the  path of the image through the lens galaxy, and this can
be used to measure the extinction, the extinction law, and the lens redshift
(e.g. Nadeau et al. 1991; Falco et al. 1999; Motta et al. 2002; Mu\~noz et
al. 2004; Mediavilla et al. 2005; Eliasdottir et al. 2006, Mosquera et al. 2011). 
The second effect, microlensing by the stars in the lens galaxy (Chang \&
Refsdal 1979), leads to wavelength dependent changes in the flux ratios because 
the effective size of the quasar accretion disk varies with wavelength
(Wisotzki et al. 1993, 1995; Wucknitz et al. 2003; Anguita et al. 2008; Bate et al. 2008;
Eigenbrod et al. 2008; Poindexter et al.2008; Floyd et al. 2009; Mosquera et al. 2009, 2011;
Blackburne et al. 2010; Mediavilla et al. 2011).  Detections of ``chromaticity''
between the images of a lensed quasar are useful for studying both phenomena
if they can be disentangled. 

In extragalactic astronomy, understanding dust is crucial to understanding 
galaxies, through its effects on estimates of star formation rates and 
galaxy evolution (e.g. Conroy et al. 2009), cosmology, through its effects on SNe~Ia 
fluxes (e.g. Jha et al. 2006), and the interpretation of gamma-ray burst afterglows
(e.g. Jakobsson et al. 2004)).  Unfortunately, classical
methods for obtaining accurate extinction curves to characterize dust cannot
be used outside the Local Group because they depend on detailed measurements
of individual stars.  Gravitational lenses are one of the best quantitative 
astrophysical probes of dust properties at intermediate redshifts given
lenses with the right amount of dust and the appropriate combinations of redshifts.
If there is too little dust, it is difficult to measure the extinction at
long wavelengths and microlensing is more likely to dominate the chromaticity.
If there is too much dust, it becomes impossible to measure extinction curves
into the rest-frame ultraviolet.  Similarly, the lens redshift must be high
enough to make the rest-frame 2175\AA\ dust feature observable, while the
source redshift must be low enough to avoid having the quasar continuum blocked
by absorption in the intergalactic medium.  Similar considerations hold for
studying chromatic microlensing over the broadest possible wavelength baseline.

We selected six lenses from the survey of extinction by Falco et al. (1999) that
roughly satisfied these criteria: HE~0512$-$3329, FBQ~0951$+$2635,
HE~1104$-$1805, H~1413$+$117, SBS~1520$+$530, and B~1600$+$434.  
As we report in \S2, we observed them in 6 filters spanning 
2200\AA\ to 8100\AA\ (F220W, F250W, F330W, F435W, F555W, F625W and
F814W) using  the Hubble Space Telescope (HST) and the ACS/HRC camera.   
This approach ensures that we can measure the image flux ratios without 
contamination from the lens or the host galaxy of the quasar.  Section 2
also outlines how we model the results to study extinction and microlensing. 
In \S3 we present the results, reporting on the extinction curves of 
three of the systems and the chromatic microlensing in one system.  
Section 4 summarizes the results and lessons for future observations.

\section{Observations and Analysis}

Table~1 provides a log of our ACS/HRC observations based on 13 HST
orbits in Cycle 12, Table~2 summarizes previous HST observations
of these systems  and Table~3 presents our new photometry.  The
observations in each filter consisted of multiple, dithered 
sub-exposures which were corrected for cosmic rays and combined
using standard methods.  We modeled the images following the
procedures of  Leh\'ar et al. (2000).  The images
were fit as a combination of point sources,  de Vaucouleurs
and exponential disk profiles convolved with TinyTim  
(Krist \& Hook 1997) PSF models.  We determined the
relative astrometry of the components and the structure of
the lens galaxy using the CASTLES H-band images where the
lens galaxy is best detected and characterized (see Fig. 1).  These
were then held fixed and the remaining images were fit
to determine the fluxes of the components in each filter. 
For the bluer filters the lens galaxy was undetected and we could easily 
confirm the model fits with  aperture photometry.

Consider multiple images $i$ of a single lensed quasar. 
Let $m_0(\lambda,t)$ be the intrinsic quasar flux at time t, expressed in 
magnitudes at observed wavelength $\lambda$. The redshifted,
extincted flux of image $i$, is then
\begin{equation}
m_i(\lambda,t)=m_0(\lambda,t) - M_i(\lambda,t) + E_i \,\,R\left(\frac{\lambda}{1+z_l}\right)
\end{equation}
where $M_i(\lambda,t)$ and $E_i=E(B-V)$ are the magnification (in magnitudes)
and extinction of image $i$, and $ R(\lambda/(1+z_l))$ is the extinction curve
at the lens redshift $z_l$. The magnification $M_i(\lambda,t)$ may depend
on wavelength and time due to microlensing effects (Wambsganss 2006 and references therein).
 By measuring the 
magnitude differences as a function of wavelength for each image pair 
(labeled $i$ and $j$), 
\begin{equation}
m_j(\lambda,t)-m_i(\lambda,t)=\Delta M(\lambda,t) +\Delta E\,\,R\left(\frac{\lambda}{1+z_l}\right)
\label{eqn:mdiff}
\end{equation}
we constrain the relative magnifications, $\Delta M(\lambda,t)=M_j(\lambda,t)-M_i(\lambda,t)$,
the extinction differences, $\Delta E(B-V)=E_j(B-V)-E_i(B-V)$, and the mean
extinction curve $R(\lambda)$. 

For the extinction law $R(\lambda)$ we used either the Cardelli \etal\ (1989; hereafter CCM) 
parametrized models for the Galactic extinction curve
or the Fitzpatrick \& Massa (1990; hereafter FM) model with its
parameters set to match the average extinction in the SMC (Gordon et al. 2003).
The main difference is that the Galactic models have a strong 2175\AA\ 
absorption feature while the SMC models do not.  One way to confirm
the presence of extinction is to estimate a dust redshift $z_d$
(Jean \& Surdej 1998, Falco \etal\ 1999), the redshift at which the 
extinction curve best fits the data, and show that it agrees with 
the observed lens redshift $z_l$.  We assume that the extinction
law is the same for all images.  Generally one image dominates
the extinction and this assumption is unimportant, but it can be
an issue if all images are significantly extincted (see Mu\~noz et al. 2004 and
McGough et al. 2005).   For models assuming there is only extinction,
we fit the data with a single $\Delta M(\lambda,t) \equiv \Delta M$,
a common differential magnification for all wavelengths which removes
any effects from the magnifications of the macro model and most of
the effects of source variability.

The second physical effect in Eqn.~\ref{eqn:mdiff} is the chromatic microlensing
produced by the $\Delta M(\lambda, t)$ term.  Because of the structure in the
microlensing magnification patterns, the changing size of the disk with wavelength
changes the magnification.  Since the observer, lens, stars and host galaxy are
all in relative motion, this magnification then changes with time.  In our
present study we will examine this using simulations.  Based on the properties
of the macro models for the lens geometry, we generate magnification patterns
using the approach of Mediavilla et al. (2006),
assuming that fraction $\alpha=0.1$ of the surface density is in stars
(i.e. that the surface density is dark matter dominated, see Kochanek et al. 2006, Mediavilla et al. 2009,
Pooley et al. 2009, Morgan et al. 2010, Mosquera et al. 2011) and we simply use $M=1 M_\odot$ stars. 
 We then convolve
the patterns with Gaussian intensity profiles to model the quasar accretion
disk, $I(R) \propto \exp(-R^2/2 r_s^2)$
where $r_s \propto \lambda^p $ characterizes the disk size at wavelength $\lambda$.
These sizes can be rescaled to a different microlensing mass as $\sqrt{M/M_\odot}$.
We make many random trials fitting the data as a function of $r_s$ and $p$, and then use Bayesian methods to estimate the size $r_s$ and the scaling
exponent $p$ for either linear or logarithmic priors on $r_s$ and linear priors on $p$,
as explained in detail in Mediavilla et al. (2011).

The last point we note is that our data are obtained at a single epoch,
so our flux ratios are really comparing $m_0(\lambda,t)-m_0(\lambda,t + \Delta t)$
where $\Delta t $ is the time delay between the images.  This means that 
intrinsic source variability combined with the time delay between the images 
can lead to wavelength dependent changes in the flux ratios which we will
ignore by assuming that $m_0(\lambda,t)-m_0(\lambda,t + \Delta t) \equiv 0$.
Particularly in the estimates of extinction, we will see negligible effects
because the parameter $\Delta M$ for the difference in the macro model
magnifications also captures any achromatic effects from ignoring time
variability.  For most of the lenses we consider, these changes will be 
small, as can be seen from the empirical quasar variability models of
MacLeod et al. (2010).  Yonehara et al. (2008) based on the ensemble SDSS quasar
structure functions (Van den Berk et al. 2004; Ivezic et al. 2004) 
estimated that there would be typical shifts of $\sim 0.1$~mag in 
single epoch observations, but that the changes in colors would be 
significantly smaller because the color changes associated with quasar
variability are far smaller than the overall variability.  We can
compensate for this problem by using modestly larger uncertainties,
 but it is really only an issue for
systems with long time delays.  The worst case is HE~1104$-$1805
which has a relatively long time delay of almost 6 months.  If we 
follow the procedures of Yonehara et al. (2008), we estimate that the time
delay can produce a bias in the shortest wavelength filter (F330W)
of roughly 0.1~mag, with a potential color change between the F330W filter 
and the  H-band of only 0.05~mag.

\section{Results}

We now consider each of the systems individually.  We found  extinction
in HE~0512$-$3329, B~1600+434 and H~1413+117 and chromatic microlensing in HE~1104$-$1805.  The
remaining two systems, FBQ~0951+2635 and SBS~1520+530, did not show 
enough of a chromaticity signature  to perform a deeper analysis given only a single epoch 
of data.  In each of the analyses, it is
necessary to determine whether the filters include any broad emission
lines, because line and continuum flux ratios can be quite different (e.g. see
Mediavilla el al. 2005).
While both are equally altered by extinction, the broad emission line
regions are more spatially extended and hence far less affected by
microlensing (e.g. see Abajas et al. 2002).  Here we are restricted to photometry,
but by tracking the filter and line locations and widths we can 
determine the degree of contamination. Fig. 1 shows HST images of the 6 systems.
Note that at least half of them are relatively disky, which is not the norm for gravitational lenses.

\subsection{HE~0512$-$3329}

HE~0512$-$3329 is a two image lensed quasar with a separation of
$0\farcs65$ and a source redshift of $z_s=1.565$ (Gregg et al. 2000).
The lens redshift is not directly measured, but the presence of a
damped Lyman$\alpha$ absorber (DLA) system and associated strong
metal line absorption systems suggests that the lens is a spiral galaxy at 
$z_l=0.93$ (Gregg et al. 2000,
Wucknitz et al. 2003).  
Table~2
presents the photometry for the lens galaxy as well as the
quasar images in the CASTLES data.  While the time delays are
not measured, they will be so short given the image separation
that the single epoch flux ratios will be unaffected by intrinsic
variability.

As we see in Fig.~2, the flux ratios have a steep dependence on the
wavelength, and the slope is little changed from the earlier 
CASTLES results or the later results from 
Eliasdottir et al. (2006).  While
there are offsets between the epochs indicative of microlensing,
they show no significant wavelength dependencies.  Wucknitz
et al. (2003) show that the broad emission line flux ratios,
which should be very little affected by microlensing, show
a wavelength dependence consistent with these trends. After
correcting for extinction using the broad emission line flux
ratios, they also find an effect from microlensing. 

Figure~3 shows the result of fitting the flux ratios assuming
they are due to differential extinction.  Like Wucknitz et al. (2003),
and unlike Eliasdottir et al. (2006), we identify a weak 2175\AA\ feature. 
The Galactic CCM extinction curve fits poorly, with $\chi^2=13$ for 4 degrees
of freedom (dof), an estimated $R(V)\lesssim 0.5$ and a best fit $R(V)\simeq 0$,
a region where the model makes no sense. 
The model with the weaker feature of the mean SMC extinction law fits
far better, with $\chi^2=2.8$ for 5 dof and $\Delta E(B-V)=0.06\pm0.01$.  If
we allow the parameter responsible for the stretch of the bump ($c3$) 
in the FM extinction law to vary, we find
a best fit with $\chi^2=1.1$ for 4 dof and parameter $c3=1.7\pm0.9$
confirming the marginal detection of the bump by Wucknitz et al. (2003). 
 The interpretation of the flux ratios
as extinction and the feature as the 2175\AA\ feature seems 
robust since we obtain a dust redshift of $z_d=0.92 \pm 0.15$ 
that is in good agreement with that of the DLA and metal
line systems at $z_l=0.93$ (Gregg et al.  2000, Wucknitz et al. 2003).
Although 25\% of the CIV emission line lies in the F435W filter, we estimate that differential microlensing between the line and continuum of order $0.2$~mag would lead to a line-induced bias in the estimated continuum flux ratios of only $\sim 0.01$~mag. This is smaller than
the photometric uncertainties and cannot explain the observed shift of $\simeq 0.1$~mag.

\subsection{B~1600+434}

B~1600+434 is a two image system with a separation of $1\farcs4$, a
source redshift of $z_s=1.59$ and a lens redshift of $z_l=0.41$
(Jackson et al. 1995) where the lens is a nearly edge on
spiral  (Jaunsen \& Hjorth 1997).  The time delay is relatively
short ($\sim 47$~days, Koopmans et al. 2000, Burud et al. 2000),
so single epoch flux ratios will be little affected by intrinsic
variability.   Not surprisingly, it was quickly found that the
image passing through the disk of the galaxy suffered from 
extinction (Jaunsen \& Hjorth 1997, Falco et al. 1999, Burud
et al. 2000).

Figure~4 shows the magnitude differences as a function of wavelength,
where the redder image A is the image passing through the disk of
the lens.  The slope of the differences is little changed from the
CASTLES observations, but there is an offset of approximately
0.2~mag.  Thus, as for HE~0512$-$3329, the dominant effect is differential
extinction with weaker effects due to microlensing that show no obvious
wavelength dependence.  Figure~5 shows a fit to the flux ratios assuming
they are due to extinction, where we have used the 6~cm radio flux ratio
(Koopmans et al. 2000) as an extinction-free anchor for the ratios.
The data are well fit by a CCM extinction law with $R(V) = 1.5\pm0.3$ and 
$\Delta E(B-V)= 0.39\pm0.02$.  The structure of the extinction law
is not tightly constrained because the 2175\AA\ feature is not only 
bluewards of our shortest wavelength filter but also lies on top 
of the Ly$\alpha$ line of the quasar.  Dai \& Kochanek (2005) 
estimated a gas column density difference between the images of
$\sim3 \times 10^{21}$~cm$^{-2}$ based on differences in the X-ray
spectra of the two images.  Using the extinction estimate of
$\Delta E(B-V) \simeq 0.1$  from Falco et al. (1999), this implied
a dust-to-gas ratio that was somewhat high.  However, if we adopt
our new estimate, we find a dust-to-gas ratio of 
$\sim 7 \times 10^{21}$~mag$^{-1}$~cm$^{-2}$ that is
very close to the typical Galactic value of 
$5.8 \times 10^{21}$~mag$^{-1}$~cm$^{-2}$ (Bohlin et al. 1978).

\subsection{H~1413+117}

H~1413+117 is a four image system with a maximum separation of $\sim
1\farcs1$ and $z_s=2.55$ (Magain et al. 1988).  The lens galaxy
is marginally detected at H-band and its redshift is unknown,
although Kneib et al. (1998) propose $z\sim 0.9$ based on the
photometric redshifts of nearby galaxies. Figure~6 shows the
magnitude differences for our HRC observations, the CASTLES data,
the Turnshek (1997) and  Chae (2001)
data, and the mid-infrared ($11\mu$m) flux ratios
from MacLeod et al. (2009). The small shifts between the 
epochs appear to be due to changes in the fluxes of images
A and D, at  levels of approximately 0.1 and 0.05~mag.
The largest wavelength dependencies correspond
to images A and B and the lack of significant changes in the colors with time
indicates that they should be attributed to extinction. The lack of a wavelength dependence 
between images D and C suggests they are little affected
by either extinction or chromatic microlensing at these wavelengths.  
The small bump in the F435W magnitude differences
including image D  (see Fig. 6) is probably due to the contamination
by the Ly${\alpha}$ emission line. In Figure 6 we also see that the flux ratios excluding image D
($m_A-m_C$, $m_B-m_C$) extend naturally into the mid-IR as might be expected for extinction,
while the flux rations including image D ($m_B-m_D$, $m_D-m_C$) show significant shifts 
($\sim 0.2$ mag) going from near-IR to the mid-IR. This seems more easily explained by
microlensing, where the near-IR emission is from the accretion disk while the mid-IR emission
is from thermal dust emission on larger scales.

We conclude that A and B images are significantly affected by differential
extinction from the lens galaxy. Unfortunately the lack of a candidate for the 2175 \AA\ feature combined 
with the unknown redshift of the lens galaxy makes it difficult to analyze the extinction.
If the bump feature is present in the lens galaxy, its absence in our observations
implies a very low lens redshift ($\lesssim 0.3$), as illustrated by the example for a lens at $z_l=0.25$ 
shown in Fig.~7.
The failure to detect the lens in the V and I-band HST observations almost certainly
guarantees that the lens redshift cannot be so low. Thus we must conclude that the extinction
law in this lens lacks a significant $2175$\AA\ feature.

\subsection{HE~1104$-$1805}

HE~1104$-$1805 is a two image lensed quasar with a relatively large separation of
$\sim 3\farcs2$ and a source redshift of $z_s=2.32$ (Wisotzki et al. 1993) and
a lens redshift of $z_l=0.73$ (Lidman et al. 2000).   Leh\'ar et al. (2000) modeled the 
system in detail using the CASTLES images.   The time delay is relatively
long ($\sim 162$~days, Morgan et al. 2008), but based on the statistics
of quasar variability discussed in \S2 our single epoch flux ratios should
not be strongly biased.  Falco et al. (1999) modeled the flux ratios as
extinction, although the X-ray absorption study by Dai et al. (2006) found
negligible differential absorption.  In fact, it was also clear from the
later light curves (Schechter et al. 2003, Poindexter et al. 2007) that 
there was significant chromatic microlensing in this system.  Indeed, as
Poindexter et al. (2008) noted in their detailed study of microlensing 
the relative colors of the two images reversed over the 
period from its discovery, very different from the limited color changes
seen in the first three lenses we considered.  Further evidence against significant extinction
is that the mid-IR flux ratios from Poindexter et al. (2007) agree well
with the emission line flux ratios (Wisotzki et al. 1993).
Figure 8 shows the magnitude differences for images A and B for each ACS/HRC filter, along with the
CASTLES magnitude differences and the  mid-IR flux ratios from Poindexter et al. (2007). We can see 
again the change in slope of the wavelength dependence between the two epochs indicating the detection of chromatic microlensing.

We separately modeled the two epochs of HST observations
using the procedures from Mediavilla et al. (2011) and Mosquera et al. (2011),
as briefly outlined in \S2, to compare the
results from single epoch models to the more complex light curve modeling
procedures used by Poindexter et al. (2008).  
Figure 9 shows the estimates
for the scale radius $r_s$ in the F336W filter (1013\AA\ in the rest frame)
and the logarithmic slope $p$ of the size with wavelength, $r_s \propto \lambda^p$, 
for the HRC data,
the CASTLES data and the combination of the two assuming either a logarithmic
or a linear prior on $r_s$.  In thin disk theory, where the disk temperature
profile is $T \propto R^{-3/4}$, we would expect to find $p=4/3$.  
Given the nature of the chromatic microlensing detected in the HRC observations
the uncertainties are substantially greater than
the ones derived from the CASTLES data, with its steeper, monotonic variations
in the flux ratios, but
the estimates (see Table 4) agree within the uncertainties.
When we combine the two results, we find  $p=1.1\pm 0.6$ and $r_s=4^{+4}_{-2}$ light-days
for the logarithmic prior on the size, and $p=1.0\pm 0.6$ and $r_s=7\pm 4$
light-days for the linear prior.  We can compare to Poindexter et al. (2008),
who used a different disk model and normalizing wavelength, by converting
the scale lengths to the half-light radii $R_{1/2}$ of the models since
Mortonson et al. (2005) showed that different microlensing models will agree
on the half-light radius of the distribution.  We transform our $r_s$
at $\lambda=3363$\AA\ to $R_{1/2}(\lambda4311)=1.18\ (4311/3363)^p\ r_s(\lambda3363)$
at the normalizing wavelength $\lambda=4311$\AA\ used by Poindexter et al. (2008),
where their radius $r_{disk}$ corresponds to a half-light radius of $R_{1/2} = 2.49 r_\lambda4311$.
In addition  we rescale our microlens mass to $M=0.3 M_{\odot}$ from $M=1 M_{\odot}$ ($r_s\propto \sqrt{M}$)
as this is closer to the expectation for normal stellar populations (see Poindexter et al. 2008).
Figure~10 shows that our combined results are in excellent agreement with those of Poindexter et al. (2008).

\subsection{FBQ~0951+2635}

FBQ~0951+2635 is a two image lens with an image separation of $ 1\farcs1$,
a source redshift of $z_s=1.24$ (Schechter et al. 1998), and a
lens redshift of $z_l=0.260\pm0.002$ (Eigenbrod et al. 2007).  The time delay is short, $\sim 16$~days
(Jakobsson et al. 2005) and several studies have detected microlensing 
variability at the level of 0.04~mag/year (e.g., Schechter et al. 1998,
Jakobsson et al. 2005, Paraficz et al. 2006).  
Figure 11 shows our 
measurements of the magnitude differences between the two images along
with the differences found by CASTLES.  The magnitudes of the
differences are smaller than in the previous four systems, and it is
clear that there is little differential extinction.  This agrees with the similar
conclusion of Mosquera et al. (2011) based on ground-based narrow band
imaging.  The UV data from HST allow to us cover the wavelength range where 
the bump feature is expected given the measured redshift of the lens galaxy.
Although we see a small feature in the F250W filter, we cannot simply attribute it to 
the 2175\AA\ bump  because it also
overlaps the Ly$\alpha$ emission line of the quasar.  The differences between the
present data and the CASTLES observations  indicate the presence of
chromatic microlensing, but the amplitudes are too small for single epoch
microlensing models to yield significant results.  Nonetheless, 
we confirm that FBQ~0951+2635 is a good candidate for future measurements, 
reinforced by the fact that
Morgan et al. (2010) were already able to estimate a disk size based
on microlensing in the R-band light curves of this source. 

\subsection{SBS~1520+530}

SBS~1520+530 is a two image lens with an image separation of 
$\simeq 1\farcs6$, a source redshift of $z_s=1.86$ and a lens
redshift of $z_l=0.72$ (Chavushyan et al. 1997).  
Burud et al. (2002) measured a time delay for the system 
of $\sim 130$ days, and Gaynullina et al. (2005) and Paraficz et al. (2006)
observed microlensing at a level of $\sim 0.14$ mag over 
roughly 4 years.   Morgan et al. (2010) estimated an R-band half-light
radius for the disk  based on modeling these light curves.
Unfortunately our HRC observations do not show 
a significant chromaticity signal to allow us to perform a deeper analysis.
Figure 12 shows the magnitude differences as a function of wavelength
for our HRC data and the CASTLES differences  presented in
Table~2.  The wavelength dependent trends
in the ACS/HRC observations are weak, indicating that the
differential extinction is very low. Interpreting the CASTLES data is difficult
because the V-band (F555W) flux ratio is so oddly different. 
We have inspected the CASTLES data several times and have been unable to find a systematic problem (e.g. missed cosmic rays) that would explain the discrepancies.

\section{Discussion and Summary}

The effects of both extinction and microlensing become larger as we
observe them at shorter wavelengths.   Unfortunately, the atmosphere
prevents us from observing into the ultraviolet from the ground, and
so we generally cannot observe the rest-frame 2175\AA\ region to search for
the characteristic feature of Galactic and LMC extinction curves or
to probe the hot regions near the inner edges of accretion disks.  
Here we surveyed six gravitational lenses with evidence for significant
wavelength dependent flux ratios in the extinction study of Falco 
et al. (1999) from the I-band into the UV (8100\AA\ to 2200\AA). 
Two of the lenses, FBQ~0951+2635 and SBS~1520+530, showed changes
with wavelength that were too small to yield interesting constraints.

It is not surprising given the selection method that three of the lenses show
significant evidence for differential extinction between the images.
We argue for extinction dominating over chromatic microlensing systems
based on the lack of evidence for significant time variability in the
color, although all three systems show small changes in the flux ratios
that are probably due to microlensing.  In the case of HE~0512$-$3329
we find evidence for a weak 2175\AA\ feature from the dust in the 
$z_l=0.93$ lens.
For B~1600+434 we cannot quite reach the wavelengths needed to 
quantify the presence of the 2175\AA\ feature, although a CCM extinction law
agrees with our observations, while 
the lack of a lens redshift for H~1413+117 limits our conclusions.  Both systems
contain significant differential extinction, and it is likely that
the dust in B~1600+434 has the 2175\AA\  feature and that the dust 
in H~1413+177 does not.

We clearly detect chromatic microlensing in HE~1104$-$1805.  If we 
estimate the wavelength dependent size of the accretion disk by modeling
our single epoch of data or the earlier CASTLES data, we find  compatible results. 
 If we combine the two single epoch estimates,
the combined result agrees with the multi-band light curve analyses
of Poindexter et al. (2008).  
Modeled as a Gaussian source $\exp(-r^2/2 r_s^2)$ with $r_s \propto \lambda^p$
and normalized at the observed wavelength $\lambda=3363$ \AA\  we find $r_s=4^{+4}_{-2}$ ($7\pm 4$)  and 
$p=1.1\pm 0.6$ ($1.0\pm 0.6$) for a logarithmic (linear) prior on $r_s$. 
These slopes are consistent with the expected slopes from standard thin disk theory
($T\propto R^{-1/p}$ with $p$=4/3), but the uncertainties are too large to draw a stronger conclusion.

\bigskip
\acknowledgements 
This research was supported by the European Community's Sixth Framework Marie Curie
Research Training Network Programme, Contract No. MRTN-CT-2004-505183 ``ANGLES'', and by
the Spanish Ministerio de Educaci\'{o}n y Ciencias (grants AYA2007-67342-C03-01/03
and AYA2010-21741-C03/02). J.A.M. is also supported by the Generalitat Valenciana with the
grant PROMETEO/2009/64. A.M.M. is also supported by Generalitat Valenciana, grant APOSTD/2010/030.
CSK is supported in part by NSF grant AST-1009756.
Based on observations made with the NASA/ESA Hubble Space Telescope, obtained
at the Space Telescope Science Institute, which is operated by the Association 
of Universities for Research in Astronomy, Inc., under NASA contract NAS 5-26555. 
These observations are associated with program GO-9896.


\newpage
\begin{figure}
\centerline{\psfig{figure=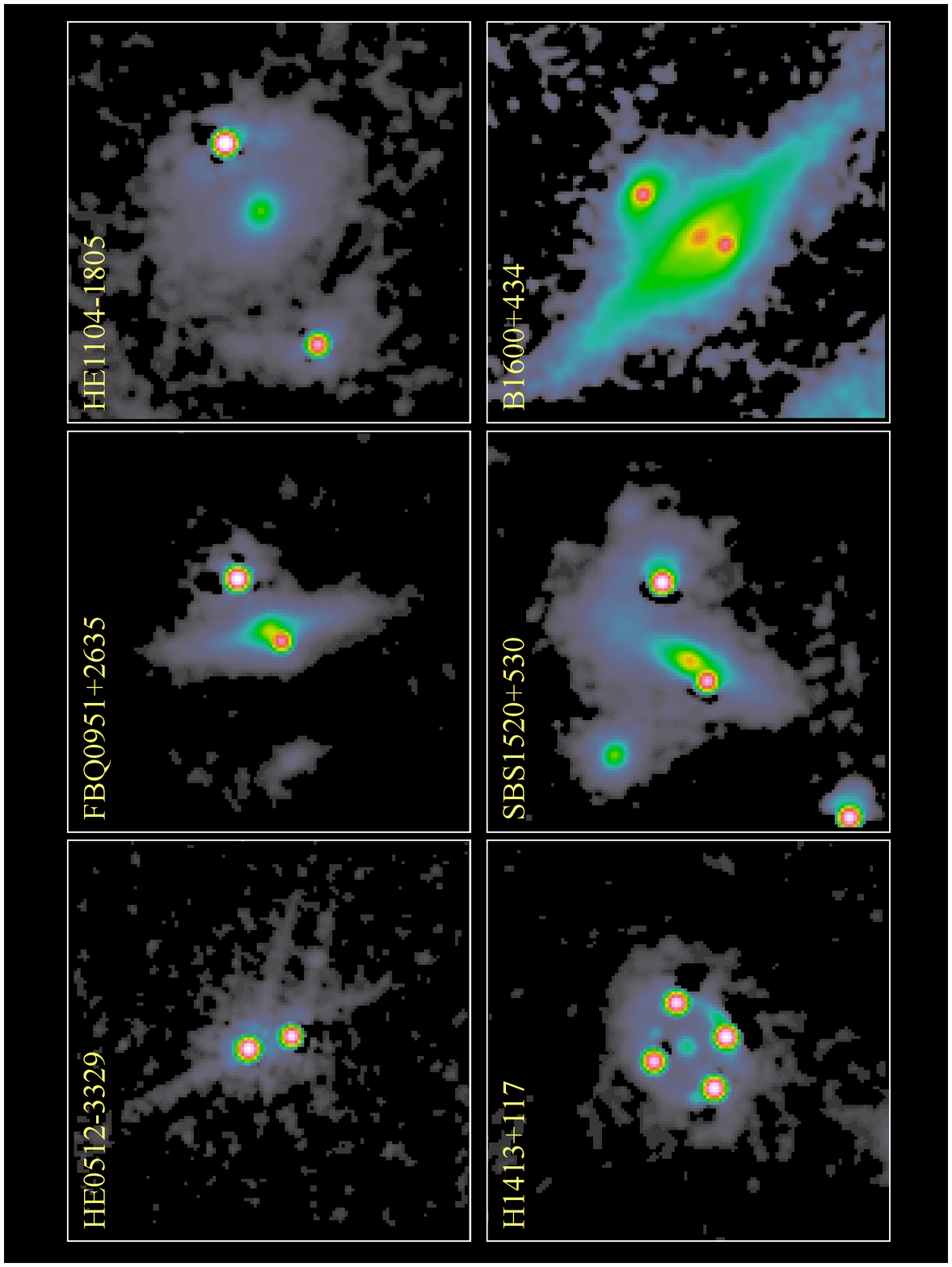,height=9.5in}}
\caption{HST NICMOS/NIC H-band  images of the six gravitational lenses in our sample.}
\end{figure}

\begin{figure}[htbp]
\centerline{\psfig{figure=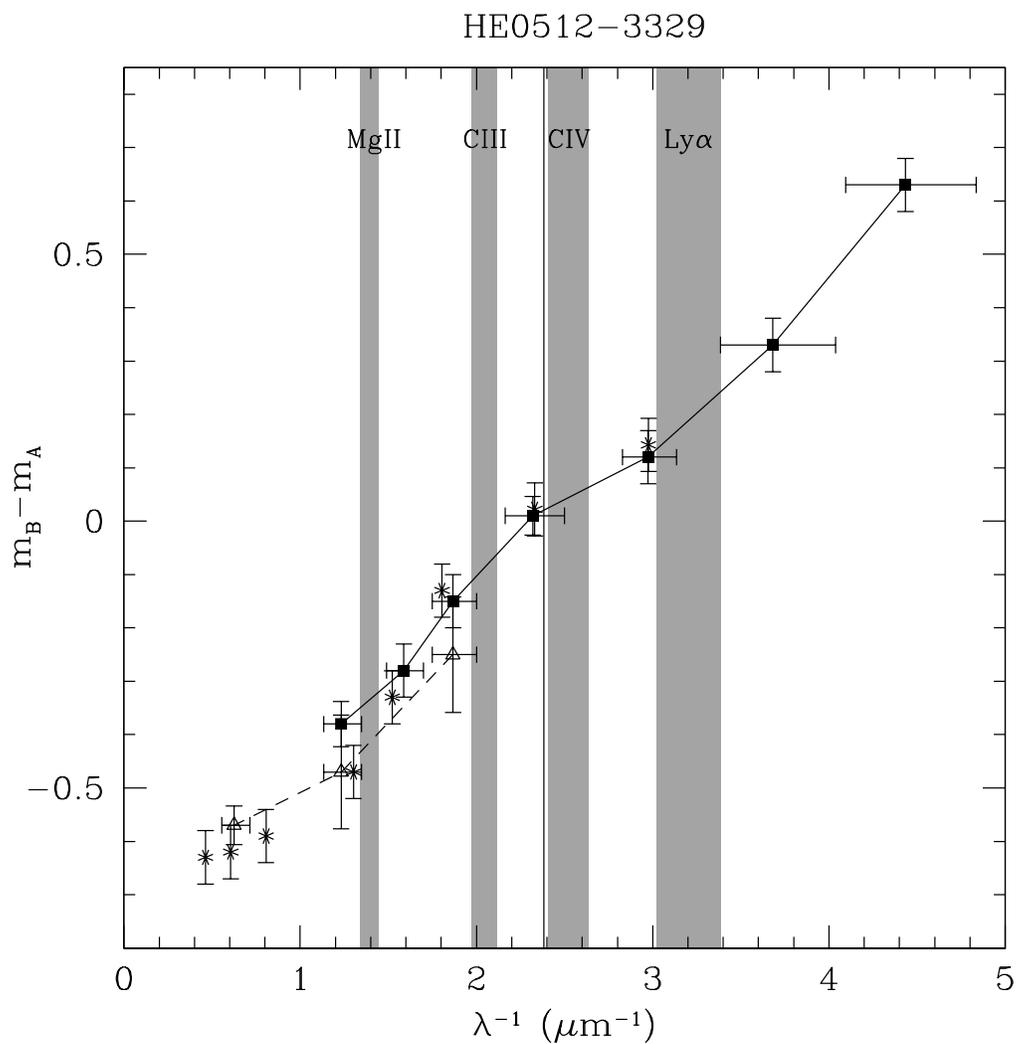,height=5.5in}}
\caption{Magnitude differences of HE~0512$-$3329 as a function of the inverse of the observed wavelength for
the new HST observations (filled squares) along with the previous CASTLES observations
(open triangles) and the ground-based data from Eliasdottir et al. (2006)  
(asterisks). The shaded regions correspond to the wavelength location and width of
the most prominent quasar broad emission lines. The vertical solid line indicates the 
expected position of the 2175\AA\ extinction curve feature based on the estimated lens redshift.
The horizontal error bars on the HST data indicate the widths of the filters.  
}
\label{default}
\end{figure}

\begin{figure}
\centerline{\psfig{figure=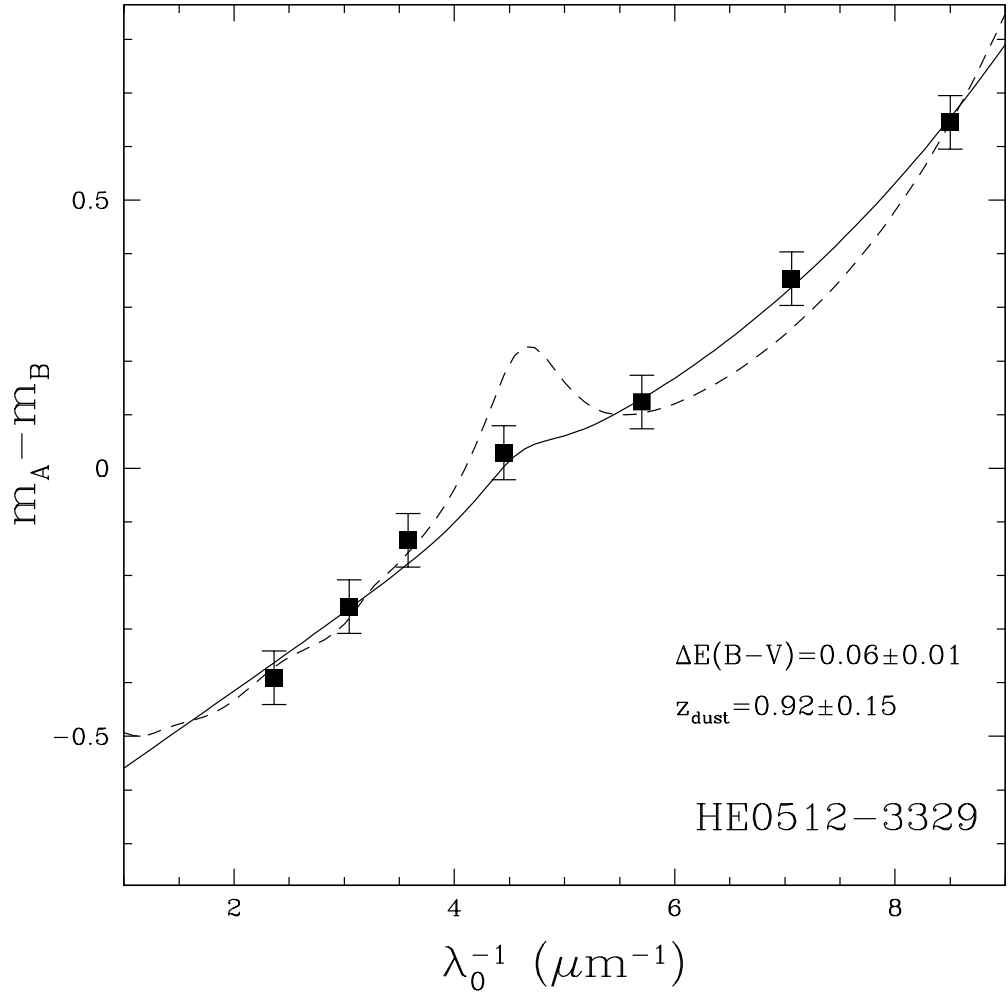,height=5.5in}}
\caption{Magnitude differences as a function of the inverse of the lens rest-frame wavelength for
the new HST observations (filled squares). The solid line shows the best fit FM extinction 
law allowing variations in the ``stretch'' of the bump and the dust redshift. The dashed line 
corresponds to the best fit for a Galactic CCM extinction law with $R(V)=0.5$.}
\end{figure}

\begin{figure}
\centerline{\psfig{figure=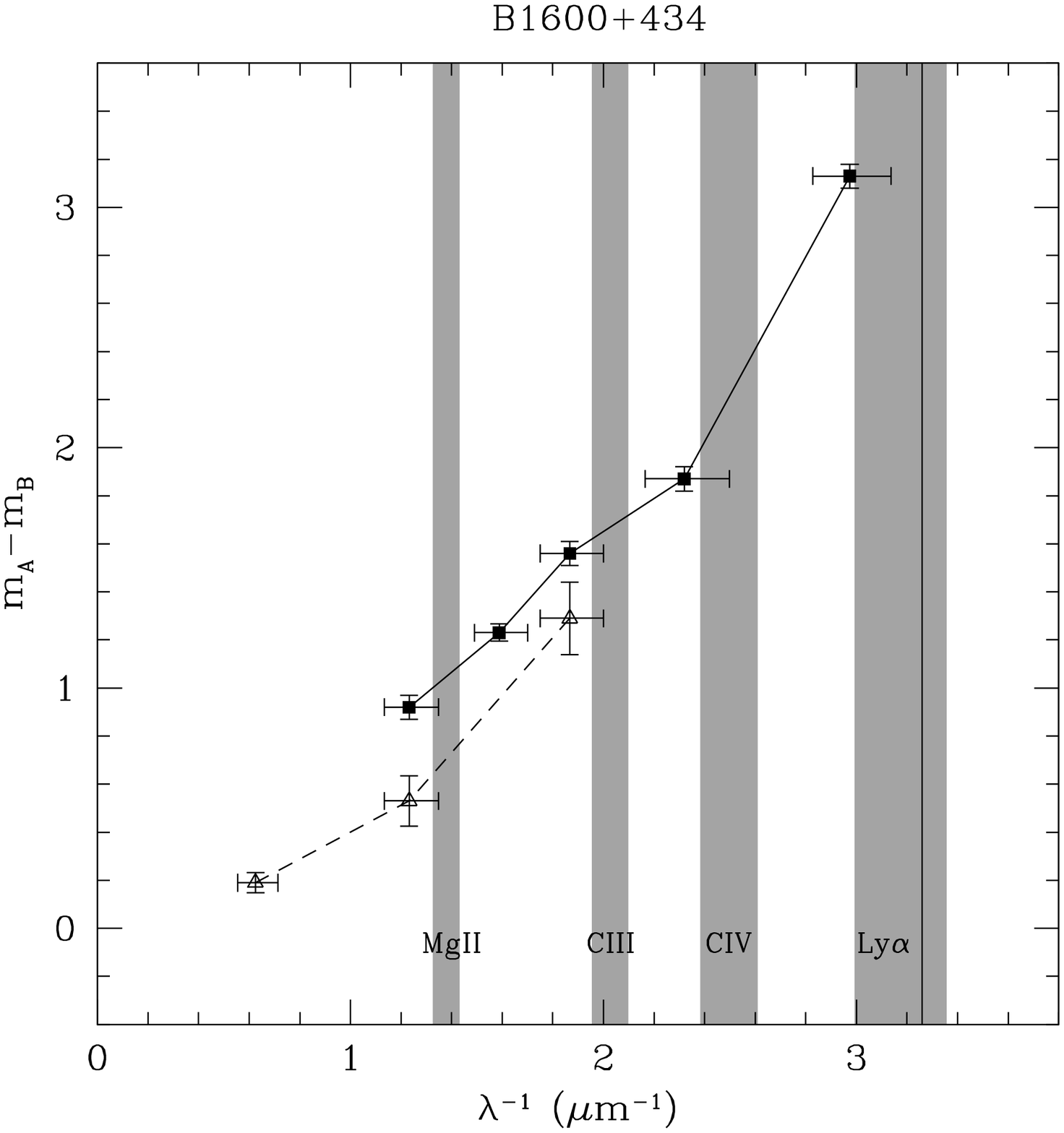,height=5.5in}}
\caption{Magnitude differences of B~1600+434 as a function of the inverse of the observed wavelength for
the new HST observations (filled squares) along with the previous CASTLES observations
(open triangles).  The format of the figure is the same as in Figure~2. }
\end{figure}

\begin{figure}
\centerline{\psfig{figure=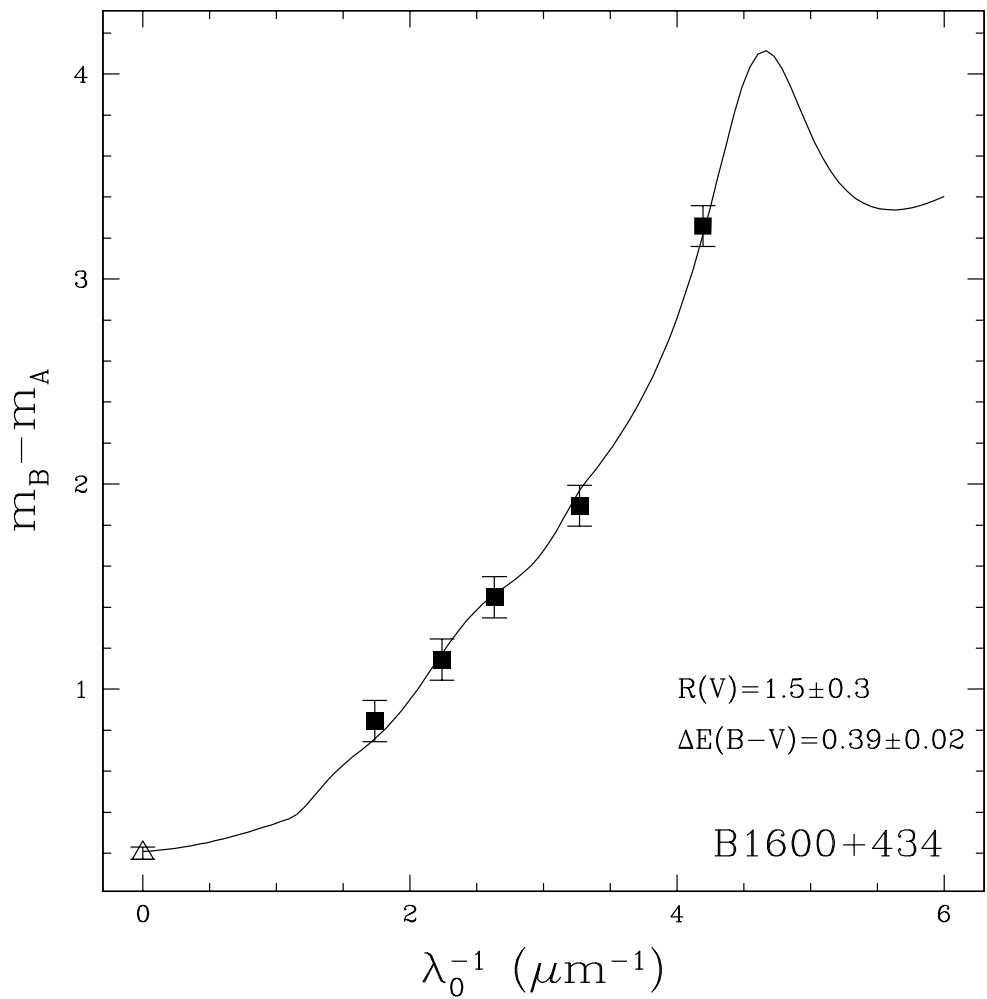,height=5.5in}}
\caption{Magnitude differences as a function of the inverse of the lens rest-frame wavelength for
the new HST observations (filled squares) and the 6~cm radio flux (open triangle) from 
Koopmans et al. (2000).  The solid line shows the best fit for a CCM extinction law. }
\end{figure}

\begin{figure}
\centerline{\psfig{figure=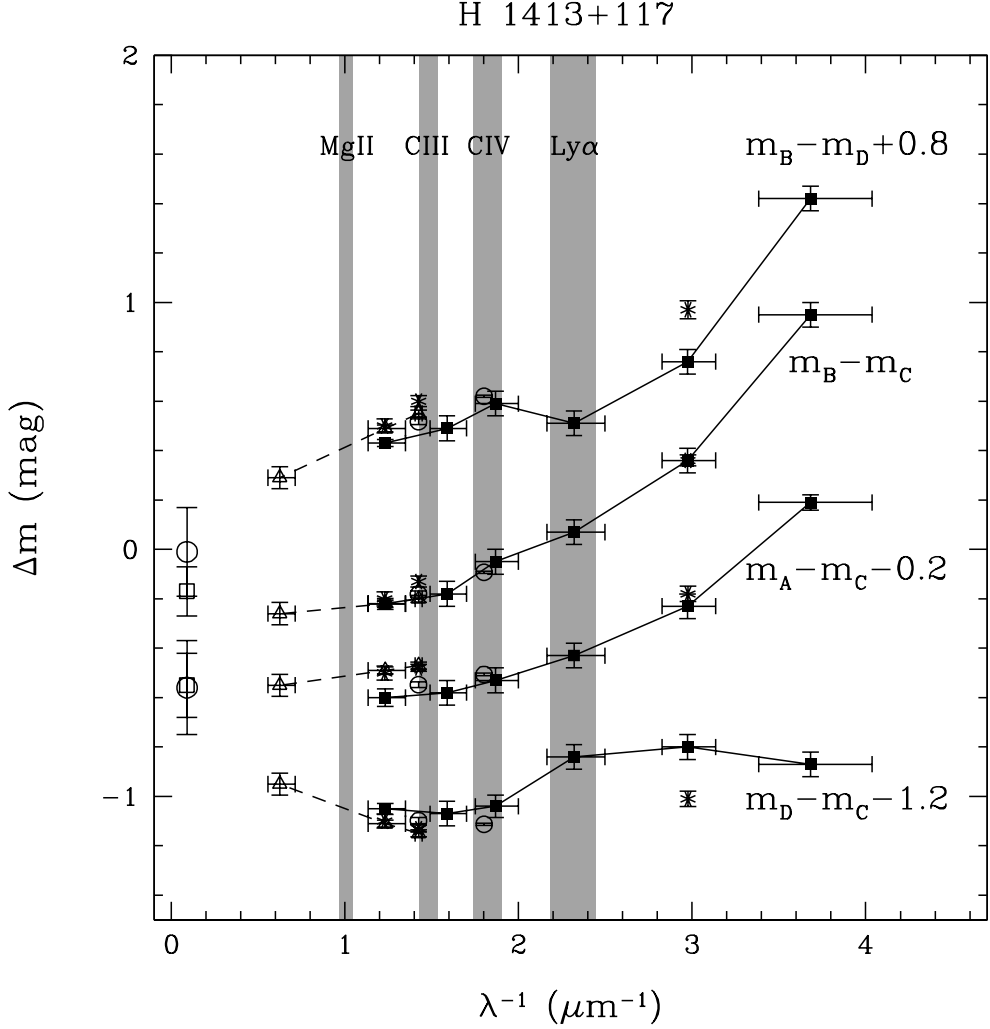,height=5.5in}}
\caption{Magnitude differences of H~1413+117 as a function of the inverse of the observed wavelength for
the new HST observations (filled squares) along with the previous CASTLES observations
(open triangles) and the mid-infrared flux (large open symbols: squares or circles when de D image is present, see text) from MacLeod et al. (2009).
We have also shown the results from Turnshek (1997) (asterisks) and Chae (2001) (open circles). In some cases
these points are completely hidden by our new measurements.
The format of the figure is the same as in Figure~2.
}
\end{figure}

\begin{figure}
\centerline{\psfig{figure=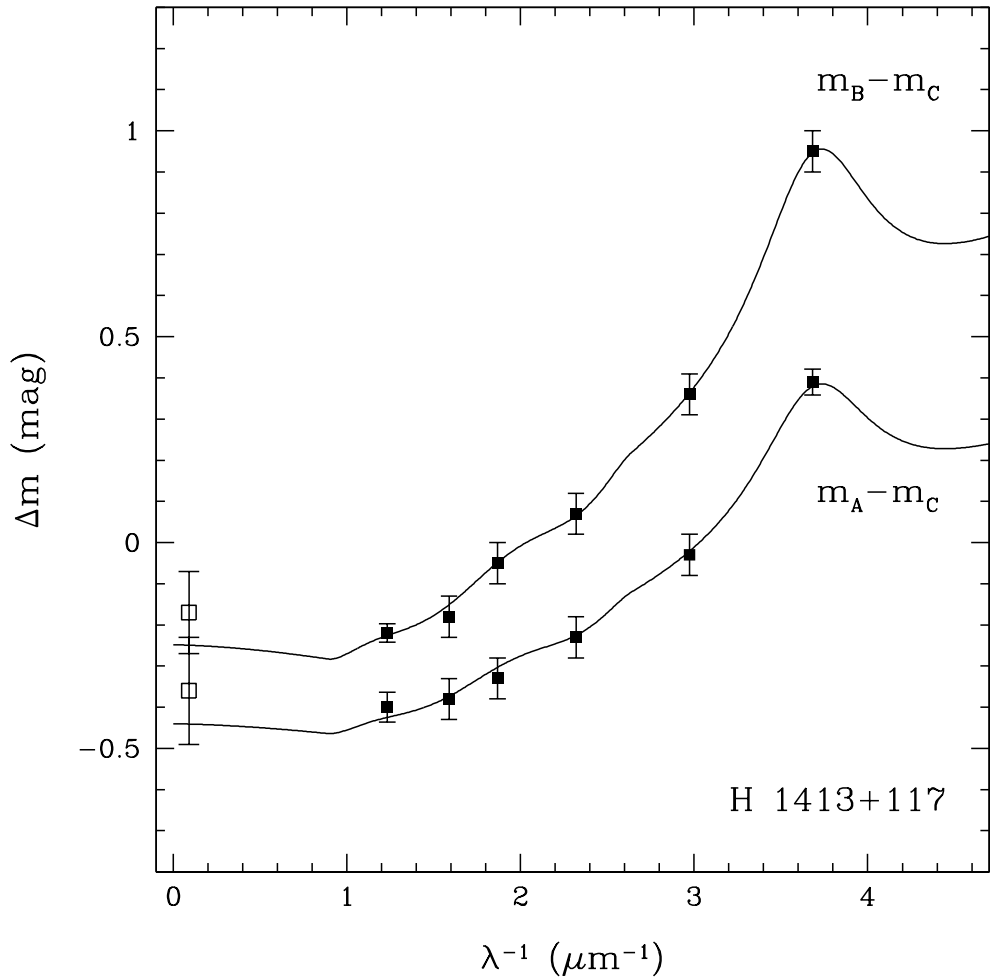,height=5.5in}}
\caption{ ACS/HRC and mid-infrared (MacLeod et al. 2009) magnitude differences $m_B-m_C$ and $m_A-m_C$,
which are strongly indicative of extinction.  As an example, the solid lines show fits with the same CCM extinction law for a fixed lens redshift $z_l=0.25$. For the more probable, higher lens redshift, where the location of the
2175\AA\ feature should be shifted to the left, successful fits require extinction laws without a strong 2175\AA\  feature.}

\end{figure}

\begin{figure}
\centerline{\psfig{figure=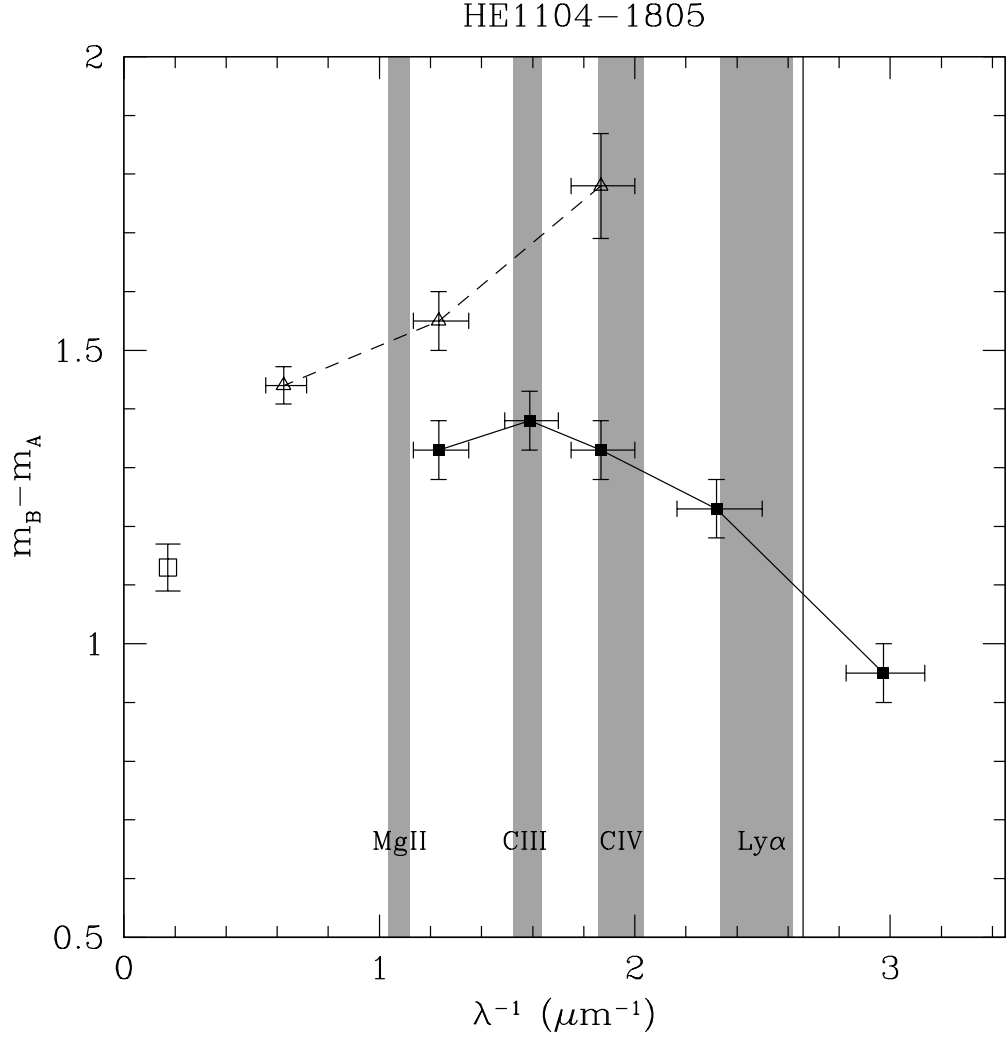,height=5.5in}}
\caption{Magnitude differences of HE~1104$-$1805 as a function of the inverse of the observed wavelength for
the new HST observations (filled squares) along with the previous CASTLES observations
(open triangles) and the mid-IR observations (open square) from Poindexter et al. (2007). 
The format of the figure is the same as in Figure~2.}
\end{figure}

\begin{figure}
\centerline{\psfig{figure=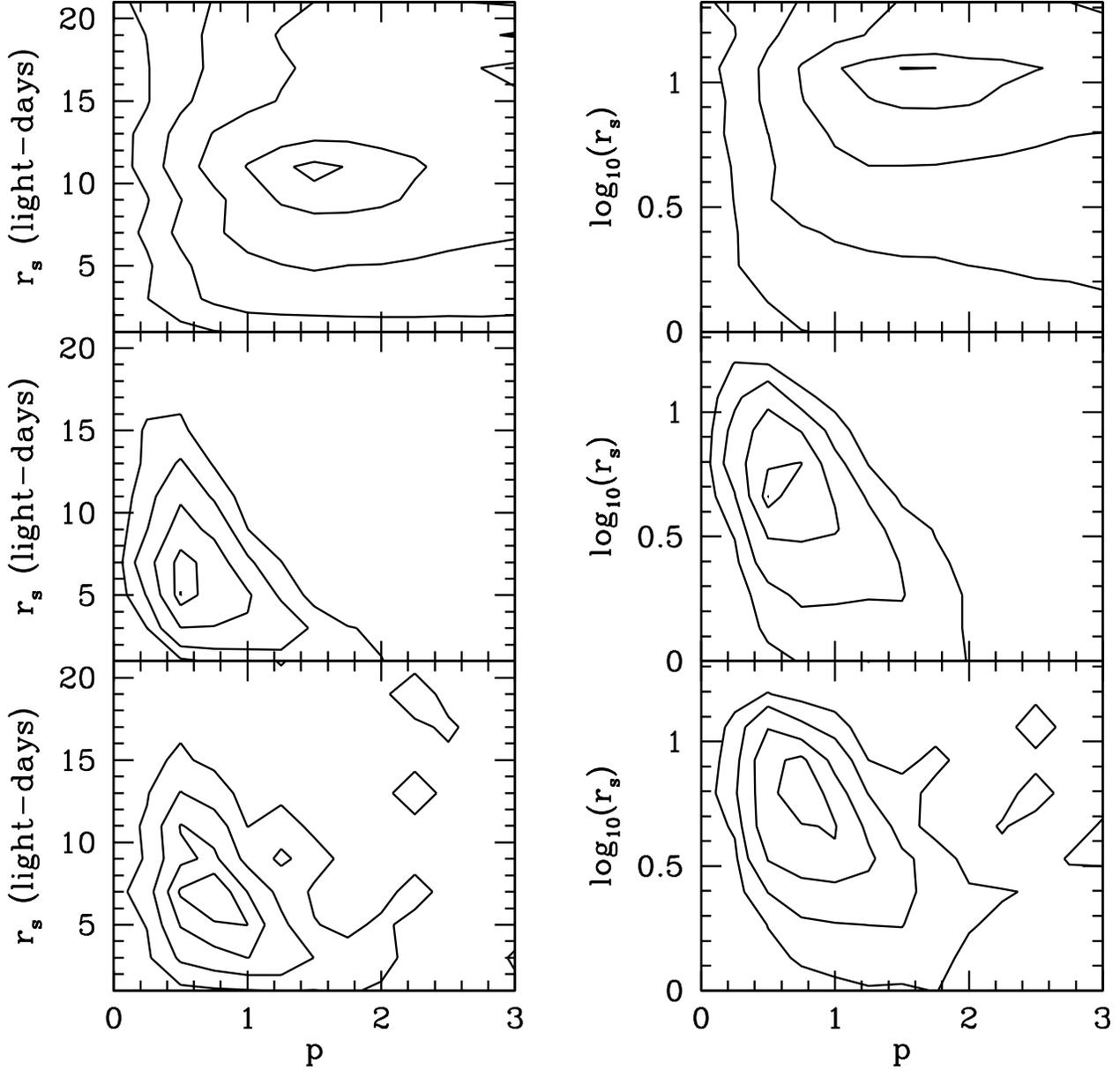,height=7in}}
\caption{Probability distributions for the size of the quasar accretion disk $r_s$ at an observed wavelength of $3363$\AA\ and the
dependence of the size on wavelength, $r_s \propto \lambda^p$, assuming a linear (left) or logarithmic (right)
prior for $r_s$, and using either the new HRC data (top), the older CASTLES data (middle) or the combined
results (bottom).  From the center, the contours are iso-probability density contours enclosing 15\%, 47\%, 68\%, and 90\% 
of the total probability, respectively. }
\end{figure}


\begin{figure}
\centerline{\psfig{figure=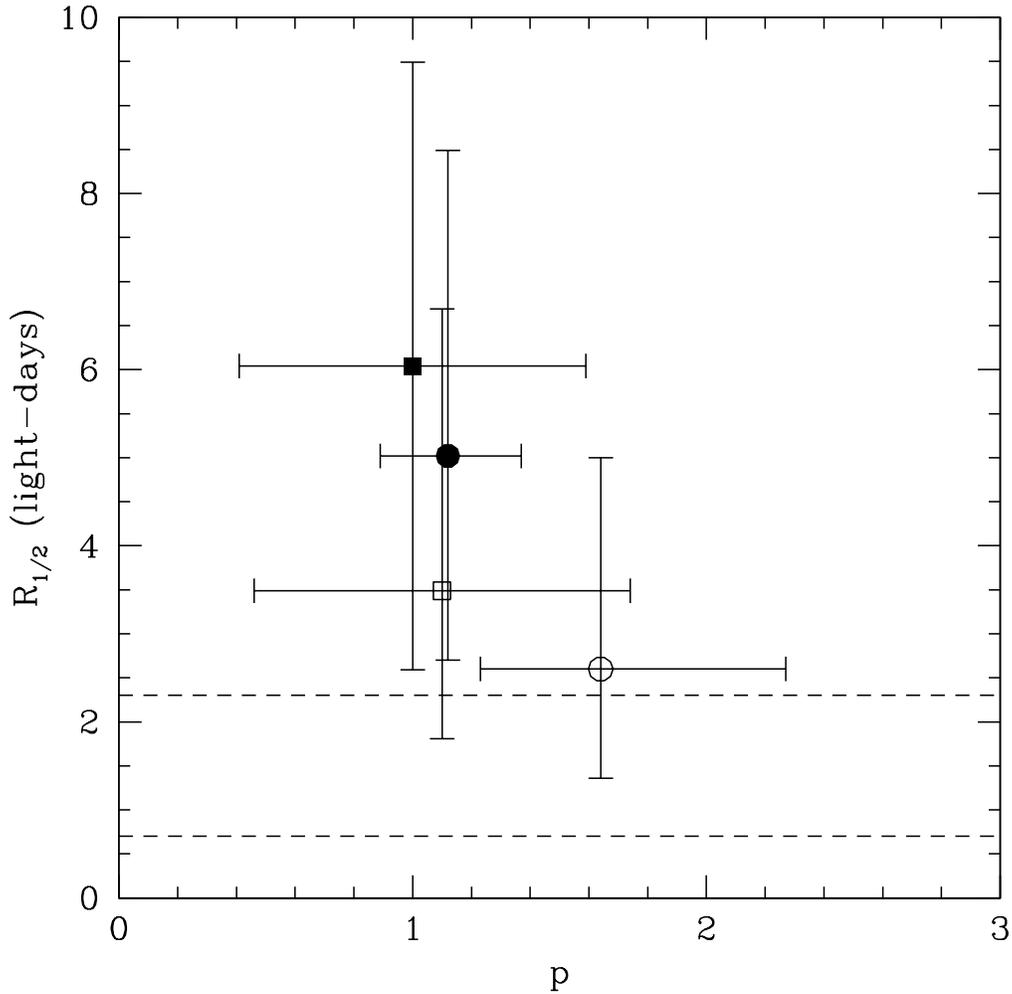,height=5.5in}}
\caption{Comparisons of the half-light radii $R_{1/2}$ at $\lambda=4311$\AA\ for our combined single epoch models
(squares) and the multi-band light curve analysis (circles) of Poindexter et al. (2008).  Open (filled) 
symbols correspond to logarithmic (linear) priors on $r_s$.  The dashed horizontal lines represent the size 
estimates inferred from the black-hole mass based on the thin disk theory (upper) or the observed I-band flux 
(lower) (see Poindexter et al. 2008). In this composition we have shifted the mean microlens mass to 0.3 M$_\odot$ (see text)
 in order to better compare to Poindexter et al. (2008).
}
\end{figure}

\begin{figure}
\centerline{\psfig{figure=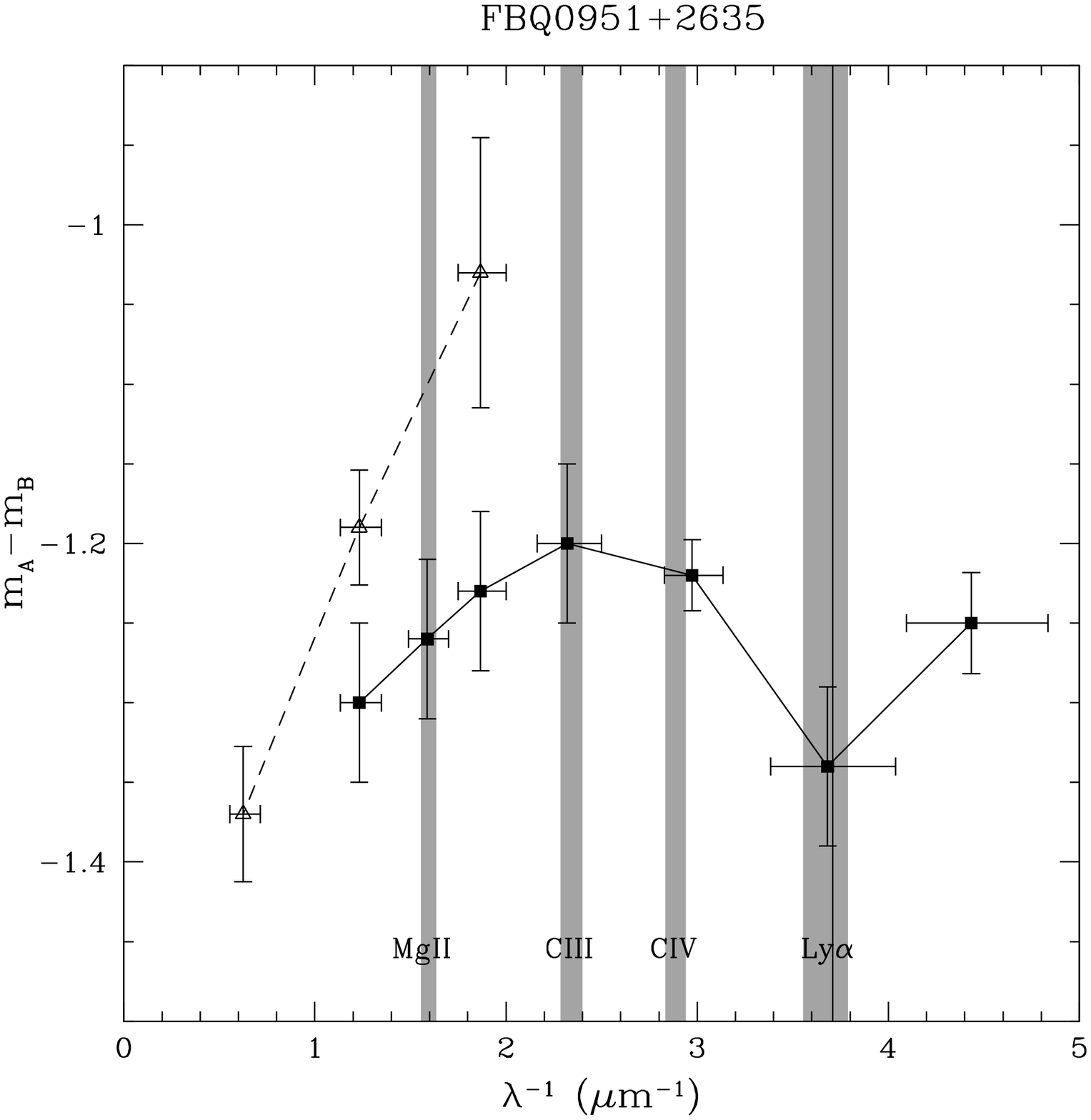,height=5.5in}}
\caption{Magnitude differences of FBQ~0951+2635 as a function of the inverse of the observed wavelength for
the new HST observations (filled squares) along with the previous CASTLES observations
(open triangles). 
The format of the figure is the same as in Figure~2.
}
\end{figure}

\begin{figure}
\centerline{\psfig{figure=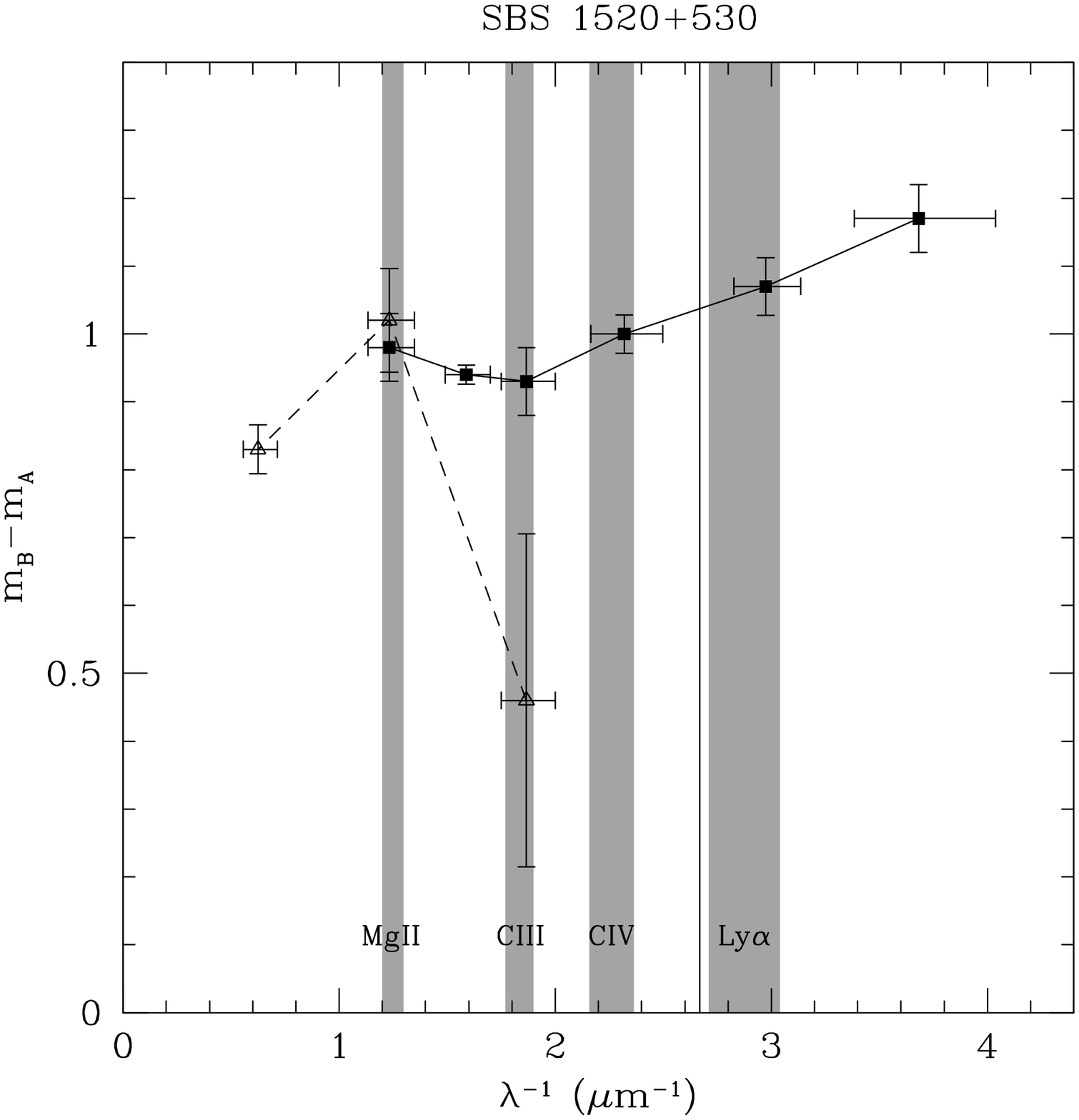,height=5.5in}}
\caption{Magnitude differences of SBS~1520+530 as a function of the inverse of the observed wavelength for
the new HST observations (filled squares) along with the previous CASTLES observations
(open triangles). 
The format of the figure is the same as in Figure~2.
}
\end{figure}

\newpage
\begin{deluxetable}{lccrcl}
\scriptsize
\tablewidth{0pt}
\tablecaption{Log of ACS/HRC Observations}
\tablehead{
\colhead{TARGET} & \colhead{DATE-OBS} & \colhead{FILTER} & \colhead{EXP} &
\colhead{No. of images} \\
       &  \colhead{(yyyy-mm-dd)}& & \colhead{(sec)}&\colhead{}\\
}
\startdata
HE~0512$-$3329    &2003-08-11    &F220W&     2136 &2\\
    &2003-08-11    &F220W&    3$\times$712 &2\\
    &2003-08-11    &F250W&      408 &2\\
    &2003-08-11    &F250W&    3$\times$136 &2\\
    &2003-08-11    &F330W&      129 &2\\
    &2003-08-11    &F330W&     3$\times$43 &2\\
    &2003-08-11    &F435W&       36 &2\\
    &2003-08-11    &F435W&     3$\times$12 &2\\
    &2003-08-11    &F555W&       27 &2\\
    &2003-08-11    &F555W&      3$\times$9 &2\\
    &2003-08-11    &F625W&       24 &2\\
    &2003-08-11    &F625W&      3$\times$8 &2\\
    &2003-08-11    &F814W&      6$\times$3 &2\\
    &2003-08-11    &F814W&       18 &2\\
    &2003-08-11    &F814W&      3$\times$6 &2\\
FBQ~0951+2635   &2003-10-06    &F220W&      368 &2\\
   &2003-10-06    &F220W&    2$\times$184 &2\\
   &2003-10-06    &F250W&      116 &2\\
   &2003-10-06    &F250W&     2$\times$58 &2\\
   &2003-10-06    &F330W&       50 &2\\
   &2003-10-06    &F330W&       25 &2\\
   &2003-10-07    &F330W&       25 &2\\
   &2003-10-06    &F435W&       16 &2\\
   &2003-10-06    &F435W&        8 &2\\
   &2003-10-07    &F435W&        8 &2\\
   &2003-10-06    &F555W&       12 &2\\
   &2003-10-06    &F555W&        6 &2\\
   &2003-10-07    &F555W&        6 &2\\
   &2003-10-06    &F625W&        8 &2\\
   &2003-10-06    &F625W&        4 &2\\
   &2003-10-07    &F625W&        4 &2\\
   &2003-10-06    &F814W&        8 &2\\
   &2003-10-06    &F814W&        4 &2\\
   &2003-10-07    &F814W&        4 &2\\
HE~1104$-$1805    &2003-11-05    &F250W&     2525 &2\\
    &2003-11-05    &F250W&      842 &2\\
    &2003-11-06    &F250W&    2$\times$842 &2\\
    &2003-11-05    &F330W&      303 &2\\
    &2003-11-05    &F330W&      303 &2\\
    &2003-11-06    &F330W&       98 &2\\
    &2003-11-06    &F330W&    2$\times$101 &2\\
    &2003-11-06    &F435W&       51 &2\\
    &2003-11-06    &F435W&     3$\times$17 &2\\
    &2003-11-06    &F555W&       30 &2\\
    &2003-11-06    &F555W&     3$\times$10 &2\\
    &2003-11-06    &F625W&       24 &2\\
    &2003-11-06    &F625W&      3$\times$8 &2\\
    &2003-11-06    &F814W&       24 &2\\
    &2003-11-06    &F814W&      3$\times$8 &2\\
H~1413+117      &2003-07-18    &F330W&      238 &4\\
      &2003-07-18    &F330W&    2$\times$119 &4\\
      &2003-07-18    &F435W&       30 &4\\
      &2003-07-18    &F435W&     2$\times$15 &4\\
      &2003-07-18    &F555W&       18 &4\\
      &2003-07-18    &F555W&      2$\times$9 &4\\
      &2003-07-18    &F625W&       16 &4\\
      &2003-07-18    &F625W&      2$\times$8 &4\\
      &2003-07-18    &F814W&       16 &4\\
      &2003-07-18    &F814W&      2$\times$8 &4\\
\tablebreak
SBS~1520+530    &2004-06-15    &F250W&     2484 &2\\
    &2004-06-15    &F250W&    3$\times$828 &2\\
    &2004-06-15    &F330W&      561 &2\\
    &2004-06-15    &F330W&    3$\times$187 &2\\
    &2004-06-15    &F435W&      141 &2\\
    &2004-06-15    &F435W&     3$\times$47 &2\\
    &2004-06-15    &F555W&       90 &2\\
    &2004-06-15    &F555W&     3$\times$30 &2\\
    &2004-06-15    &F625W&       68 &2\\
    &2004-06-15    &F625W&     3$\times$26 &2\\
    &2004-06-15    &F814W&       62 &2\\
    &2004-06-15    &F814W&     3$\times$24 &2\\
B~1600+434      &2003-08-17    &F330W&   4$\times$1080 &2\\
     &2003-08-17    &F435W&    4$\times$498 &2\\
      &2003-08-17    &F555W&    4$\times$402 &2\\
     &2003-08-17    &F625W&    4$\times$312 &2\\
      &2003-08-17    &F814W&    4$\times$293 &2\\
\enddata
\end{deluxetable}

\begin{deluxetable}{lcccccc}
\scriptsize
\tablewidth{0pt}
\tablecaption{CASTLES Photometry}
\tablehead{ 
\colhead{Lens} &\colhead{Component}  & $\Delta$R.A. ('') & $\Delta$Dec ('')& \colhead{F555W$^{\dagger}$} &\colhead{F814W} &\colhead{F160W}}
\startdata
HE~0512$-$3329 & image A &  $0.182 \pm 0.003$ & $0.621\pm 0.003$& $18.15\pm 0.06$ & $16.81\pm0.08$ & $15.81\pm0.02$\\
               & image B &  $0$ & $0$& $18.40\pm 0.09$ & $17.28\pm0.07$ & $16.38\pm0.03$\\
               & lens G &   $0.09 \pm 0.07$ & $0.37\pm 0.10$& $22.1\pm 0.6$ & $20.9\pm0.7$ & $19.1\pm0.8$\\
FBQ~0951+2635  & image A & $0$ & $0$& $17.29\pm 0.06$ & $16.70\pm0.03$ & $15.62\pm0.03$\\
               & image B & $0.900 \pm 0.003$ & $-0.635\pm 0.003$& $18.32\pm 0.06$ & $17.89\pm0.02$ & $16.99\pm0.03$\\
               & lens G &  $0.760 \pm 0.003$ & $-0.455 \pm 0.003$& $21.02\pm 0.04$ & $19.67\pm0.03$ & $17.86\pm0.14$\\
HE~1104$-$1805 & image A &  $0$ & $0$& $16.92\pm 0.06$ & $16.40\pm0.03$ & $15.91\pm0.01$\\
               & image B &  $2.901 \pm 0.003$ & $-1.332\pm 0.003$& $18.70\pm 0.08$ & $17.95\pm0.04$ & $17.35\pm0.03$\\
               & lens G &   $0.965 \pm 0.003$ & $-0.500\pm 0.003$& $23.26\pm 0.30$ & $20.01\pm0.10$ & $17.52\pm0.09$\\
H~1413+117     & image A &    $0$ & $0$& $18.00\pm 0.01$ & $17.77\pm0.01$ & $15.83\pm0.04$\\
               & image B &    $0.744 \pm 0.003$ & $0.168\pm 0.003$& $18.07\pm 0.01$ & $17.84\pm0.01$ & $15.92\pm0.03$\\
               & image C &    $-0.492 \pm 0.003$ & $0.713\pm 0.003$& $18.27\pm 0.01$ & $18.06\pm0.01$ & $16.18\pm0.02$\\
               & image D &    $0.354 \pm 0.003$ & $1.040\pm 0.003$& $18.32\pm 0.01$ & $18.15\pm0.01$ & $16.43\pm0.03$\\
               & lens G &     $0.142 \pm 0.003$ & $0.561\pm 0.003$& $-$ & $-$ & $18.61\pm0.03$\\
SBS~1520+530   & image A &  $0$ & $0$& $18.83\pm0.05$ & $17.97\pm0.03$ & $17.58\pm0.02$\\
               & image B &  $1.429 \pm 0.003$ & $-0.652\pm 0.003$& $19.29\pm 0.24$ & $18.99\pm0.07$ & $18.41\pm0.03$\\
              & lens G &   $1.141 \pm 0.003$ & $-0.395\pm 0.003$& $23.40\pm 2.00$ & $20.16\pm0.11$ & $18.22\pm0.05$\\
B~1600+434     & image A &    $0$ & $0$& $23.61\pm 0.12$ & $21.92\pm0.10$ & $20.66\pm0.03$\\
               & image B &    $-0.720\pm 0.003$ & $1.183\pm 0.004$& $22.32\pm 0.09$ & $21.39\pm0.03$ & $20.47\pm0.03$\\
               & lens G &     $-0.110 \pm 0.003$ & $0.369\pm 0.004$& $-$& $20.78\pm0.06$ & $18.30\pm0.13$\\
\tableline
\enddata
\tablecomments{$^{\dagger}$ For the system H~1413+117 it corresponds to the filter F702W}
\label{systems}
\end{deluxetable}

\begin{deluxetable}{lccccccccc}
\scriptsize
\tablewidth{0pt}
\tablecaption{ACS/HRC Photometry}
\tablehead{ 
\colhead{Lens} &\colhead{Image} & \colhead{F220W} & \colhead{F250W} &\colhead{F330W} 
&\colhead{F435W} & \colhead{F555W} & \colhead{F625W} &\colhead{F814W} }
\startdata
HE~0512$-$3329 & A & 18.96$\pm$0.11& 18.07$\pm$0.23& 17.67$\pm$0.13& 18.67$\pm$0.03& 18.10$\pm$0.05& 17.60$\pm$0.05& 16.98$\pm$0.03  \\
               & B & 18.33$\pm$0.04& 17.74$\pm$0.02& 17.55$\pm$0.03& 18.66$\pm$0.02& 18.25$\pm$0.04& 17.88$\pm$0.03& 17.36$\pm$0.03 \\
FBQ~0951+2635  & A & 16.72$\pm$0.01& 16.36$\pm$0.03& 16.60$\pm$0.02& 17.80$\pm$0.04& 17.48$\pm$0.03& 17.14$\pm$0.03& 16.82$\pm$0.03 \\ 
               & B & 17.97$\pm$0.03& 17.70$\pm$0.08& 17.82$\pm$0.01& 19.00$\pm$0.08& 18.71$\pm$0.04& 18.40$\pm$0.10& 18.12$\pm$0.05 \\
HE~1104$-$1805 & A & \nodata & \nodata & 17.25$\pm$0.05& 17.81$\pm$0.07& 17.57$\pm$0.10& 17.33$\pm$0.06& 16.85$\pm$0.05 \\
               & B & \nodata & \nodata & 18.20$\pm$0.09& 19.04$\pm$0.12& 18.90$\pm$0.11& 18.71$\pm$0.12& 18.18$\pm$0.07 \\
H~1413+117     & A & \nodata & 20.84$\pm$0.03& 18.97$\pm$0.07& 18.61$\pm$0.06& 18.20$\pm$0.09& 17.75$\pm$0.02& 17.70$\pm$0.03 \\
               & B & \nodata & 21.40$\pm$0.12& 19.36$\pm$0.08& 18.91$\pm$0.12& 18.48$\pm$0.07& 17.95$\pm$0.05& 17.88$\pm$0.01 \\
               & C & \nodata & 20.45$\pm$0.01& 19.00$\pm$0.01& 18.84$\pm$0.03& 18.53$\pm$0.04& 18.13$\pm$0.05& 18.10$\pm$0.02 \\
               & D & \nodata & 20.78$\pm$0.10& 19.40$\pm$0.06& 19.20$\pm$0.09& 18.69$\pm$0.02& 18.26$\pm$0.04& 18.25$\pm$0.01 \\
SBS~1520+530   & A & \nodata & 18.23$\pm$0.07& 17.86$\pm$0.03& 18.94$\pm$0.02& 18.73$\pm$0.03& 18.52$\pm$0.01& 18.12$\pm$0.04 \\
               & B & \nodata & 19.40$\pm$0.10& 18.93$\pm$0.03& 19.94$\pm$0.02& 19.66$\pm$0.04& 19.46$\pm$0.01& 19.10$\pm$0.04 \\
B~1600+434     & A & \nodata & \nodata & 25.68$\pm$0.47& 25.36$\pm$0.11& 24.63$\pm$0.10& 23.67$\pm$0.02& 22.68$\pm$0.03 \\
               & B & \nodata & \nodata & 22.55$\pm$0.17& 23.49$\pm$0.25& 23.07$\pm$0.11& 22.44$\pm$0.03& 21.76$\pm$0.06 \\
\tableline
\enddata
\label{photo}
\end{deluxetable}

\begin{deluxetable}{lccc}
\tablewidth{0pt}
\tablecaption{Quasar Accretion Disk Measurements for HE~1104$-$1805}
\tablehead{ 
& \colhead{ACS} &\colhead{CASTLES} & \colhead{ACS x CASTLES}}
\startdata
 & & Logarithmic prior & \\
\tableline
$r_s$ \small{(light-days)} & 6$^{+8}_{-4}$ & 4$^{+4}_{-2}$ & 4$^{+4}_{-2}$ \\
$p $  & 1.8$\pm$0.8 & 1.0$\pm$0.7 & 1.1$\pm$0.6   \\
\tableline
 & & Linear prior & \\   
\tableline
$r_s$ \small{(light-days)}  & 12$\pm$6 & 7$\pm$4 & 7$\pm$4 \\
$p $  & 1.8$\pm$0.8 & 0.9$\pm$0.6 & 1.0$\pm$0.6   \\
\tableline
\enddata
\tablecomments{ $r_s$ is the size
of the quasar accretion disk modeled as a Gaussian ($I(R)\propto \exp(-R^2/2r_s^2)$)
at the observed wavelength $\lambda=3363\,\rm \AA$ and $p$ is the power
law of the size variation with wavelength ($r_s(\lambda)\propto \lambda^p$).}
\label{photo}
\end{deluxetable}

\end{document}